\documentclass[aps,prb,10pt,twocolumn,floatfix,showpacs,superscriptaddress]{revtex4-1}
\usepackage[colorlinks=true,citecolor=blue,linkcolor=blue,urlcolor=blue]{hyperref}
\usepackage{amsmath}
\usepackage{amssymb}
\usepackage{graphicx,amsmath,amsfonts,bm}
\usepackage{epstopdf}
\usepackage{bm}
\usepackage{bbm}
\usepackage{color}
\usepackage{relsize}
\usepackage{dsfont}
\usepackage{braket}
\usepackage[caption=false]{subfig}
\usepackage{hyperref}
\usepackage[all]{hypcap} 
\usepackage[T1]{fontenc}
\usepackage{dsfont}
\usepackage{mathbbol}
\usepackage{wasysym}
\usepackage{physics}

\graphicspath{{Pictures/},{Pictures3/}}

\begin{document}
\title{Effects of spin-orbit coupling in a valley chiral kagome network} 
\author{P. Wittig}
\affiliation{Institut f\"ur Mathematische Physik, Technische Universit\"at Braunschweig, D-38106 Braunschweig, Germany} 
\author{F. Dominguez}
\affiliation{Institut f\"ur Mathematische Physik, Technische Universit\"at Braunschweig, D-38106 Braunschweig, Germany} 
\author{P. Recher}
\affiliation{Institut f\"ur Mathematische Physik, Technische Universit\"at Braunschweig, D-38106 Braunschweig, Germany} 
\affiliation{Laboratory of Emerging Nanometrology, D-38106 Braunschweig, Germany}
\date{June 26, 2024}

\begin{abstract}
Valley chiral kagome networks can arise in various situations like, for example, in double-aligned graphene-hexagonal boron nitride and periodically strained graphene.
Here, we construct a phenomenological scattering model based on the symmetries of the network to investigate the energy spectrum and magnetotransport in this system.   
Additionally, we consider the effects of a finite Rashba spin-orbit coupling on the transport properties of the kagome network. We identify conditions where the interplay of the Rashba spin-orbit coupling and the geometry of the lattice results in a reduction of the periodicity of the magnetoconductance and characteristic sharp resonances. Moreover, we find a finite spin polarization of the conductance, which could be exploited in spintronic devices.
\end{abstract}
\maketitle

\section{Introduction}
Twistronics \cite{Hennighausen2021} has introduced a new route to manipulate the band structure of crystalline systems, yielding the emergence of correlated phases by selectively altering the energy dispersion of the bands. The first realization of twistronics was found in twisted bilayer graphene, where close to the so-called magic angle correlated phases arise, such as superconducting, strange metal, and Mott insulating phases \cite{Kim2017,Po2018,Cao2018,Cao2018a,Lu2019,Xie2019,Yankowitz2019,Sharpe2019,Kerelsky2019,Choi2019,Cao2020,Zondiner2020,Wong2020}.
Also, it has opened the possibility to design new phases of matter, like the Chern mosaic system, where the local valley Chern number is changing within different regions of the bulk, leading to the emergence of a network of valley chiral modes propagating inside the material. An example of this phase was predicted in minimally twisted bilayer graphene in the presence of an interlayer bias \cite{Prada2013, Zhang2013}. There, a triangular Chern mosaic arises with a valley Chern number difference of $\pm2$ between different plackets, yielding a triangular network with two valley chiral modes propagating in the bulk. This system has been analyzed theoretically \cite{Prada2013,Zhang2013, Xu2019,Wu2019,Walet_2019,Chou2020,Tsim2020,Hou2020,Fleischmann2020,Koenig2020,Chen2020,Chou2021,Vakhtel2022a,park2023network} and measured experimentally \cite{Ju2015,Yin2016,Huang2018,Sunku2018,Rickhaus2018,Verbakel2021,Mahapatra22}.
Chern mosaic systems are not unique to twisted bilayer graphene; they  have been also found in trilayer graphene \cite{devakul2023magicangle}, double-aligned graphene-hexagonal boron nitride \cite{Vladimir2022Kagome, devries2023kagome}, and periodically strained graphene \cite{DeBeule22Kagome}. In these two latter cases, instead of having a triangular lattice structure it exhibits a kagome  lattice structure and a single valley chiral mode propagating along the sides of the hexagons and triangles [see Fig.~\ref{fig.KagomeTriangle}(a)].

A simple way to model phenomenologically these networks consists of using the Chalker-Coddington scattering model \cite{Chalker_1988} adapted to the geometry and symmetries of the system. In this way, one can obtain the band structure and magnetotransport in a straightforward way \cite{Efimkin2018,DeBeule2020,Wittig2023}. Indeed, it also allows to study topology \cite{DeBeule2021Floquet}, interactions \cite{Wu2019,Chou2019,Chen2020, Koenig2020, Chou2021, park2023network}, effective Bloch oscillations \cite{Vakhtel2022a} or multiterminal transport \cite{DeBeule2021}. Here, we follow the same principles and set up a phenomenological scattering model to study the energy spectrum and the transport properties of the valley chiral kagome network. Additionally, we consider the presence of Rashba spin-orbit coupling, which could be interesting for spintronic applications\cite{Han2014,Avsar2020}. Although the primary examples of such networks are graphene-based systems and spin-orbit coupling in graphene itself is very small (order of $\mu$eV) \cite{Kane2005,Huertas2006,Huertas2007,Gmitra2009,Sichau2019}, one could increase it through proximity to a substrate. Of special interest are transition metal dichalcogenides, which can enhance the spin-orbit coupling up to order meV \cite{Avsar_2014,Gmitra2016,Garcia_2017,Wang2019,F_l_p_2021,Sun2023}. Motivated by these realizations, we include spin-orbit coupling into the scattering network and study its impact on the energy spectrum and magnetotransport. 

The structure of the paper is as follows: First we introduce the scattering model of the kagome network in Sec.~\ref{Sec.Model}. Also, we show how to transform the kagome network into a triangular network by combining scattering matrices. Then, we calculate the network energy spectrum and magnetotransport in Sec.~\ref{Sec.wosoi}. Next, in Sec.~\ref{Sec.SO} we add the presence of a finite spin-orbit coupling and calculate the corresponding energy spectrum in Sec.~\ref{Sec.BandsSO} and magnetotransport in Sec.~\ref{Sec.CondSO}. Finally, we analyze the spin polarization of the conductance in Sec.~\ref{Sec.Pol}.

\section{Kagome scattering network model} \label{Sec.Model}
\begin{figure}
	\includegraphics[width=0.5\textwidth]{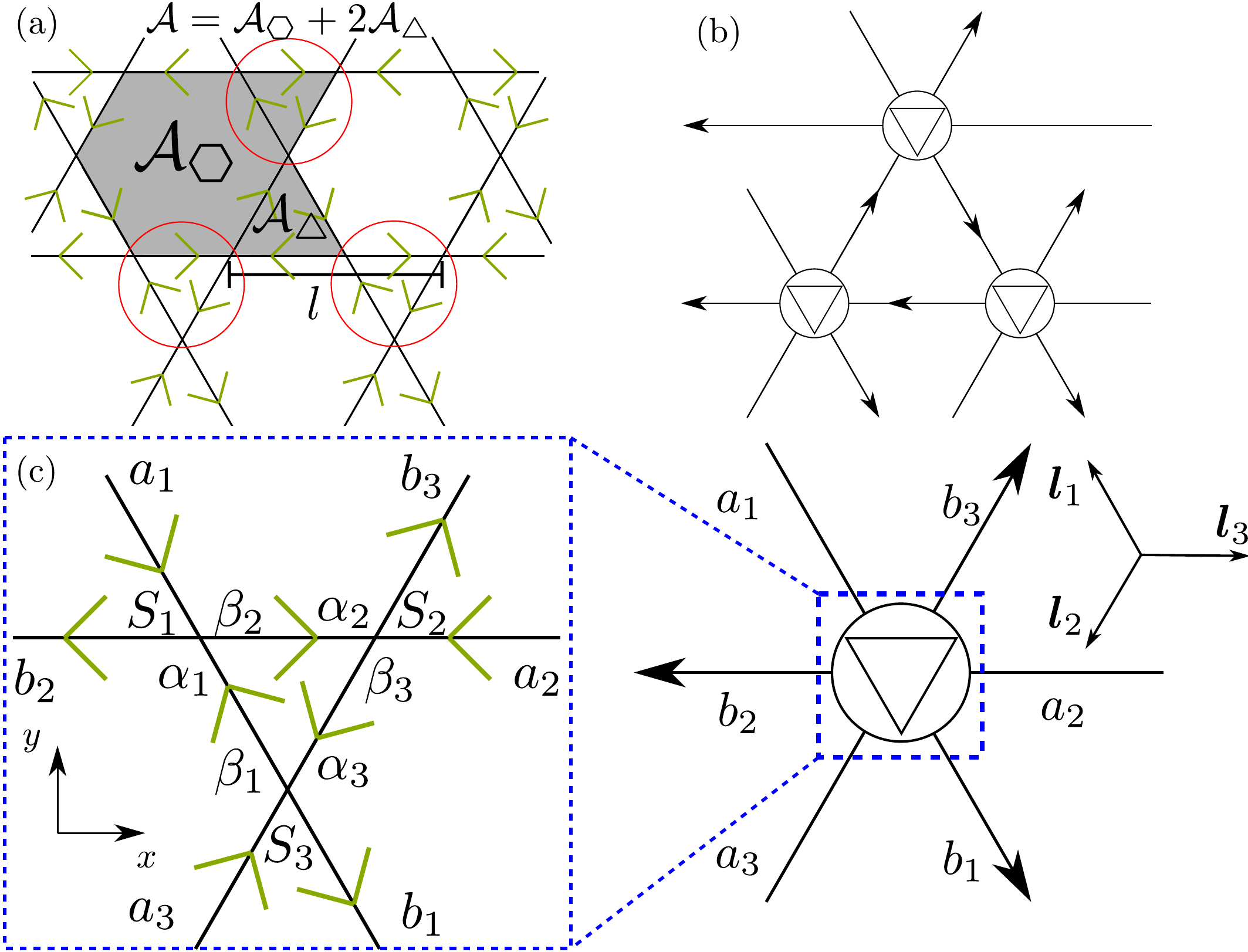}
	\caption{(a) The kagome network consists of hexagons with area $\mathcal{A}_{\hexagon}$ and triangles with area $\mathcal{A}_{\triangle}$. The unit cell is given by one hexagon and two triangles and therefore has an area of $\mathcal{A}=\mathcal{A}_{\hexagon}+2\mathcal{A}_{\triangle}$, which is the shaded area in the figure. The green arrows show the direction of the chiral modes. The characteristic length scale of the system $l$ is shown. Through combining the scattering matrices of triangles inside of the red circles the kagome scattering network is transformed to a triangular one [see (b)]. (c) Single triangle in the kagome network. Incoming modes are labeled with $a_i$ $(\alpha_i)$ outside (inside) of the triangle. Outgoing modes are labeled similarly with $b_i$ $(\beta_i)$. Also the coordinate axes in $x$ and $y$ direction and the lattice vectors $\bm{l}_i$ are depicted here.} \label{fig.KagomeTriangle}
\end{figure}

The kagome chiral network of a given valley consists of chiral modes propagating along the links of hexagons and triangles and scattering on the nodes [see in Fig.~\ref{fig.KagomeTriangle}(a)]. This system preserves time-reversal symmetry and, thus, on the opposite valley, chiral modes propagate in the opposite direction.
Here, we introduce a phenomenological scattering model based on the symmetries of the lattice ($C_3$ and $\mathcal{M}_x \mathcal{T}$\footnote{Here, $\mathcal{M}_x$ is the mirror symmetry relative to the $y$ axis [cf. inset Fig.~\ref{fig.KagomeTriangle}(c)] and $\mathcal{T}$ is time-reversal symmetry.}) to describe the propagation and scattering processes of the chiral modes. 

We start describing the scattering processes taking place at the nodes of the triangles. Using the notation depicted in Fig.~\ref{fig.KagomeTriangle}(c), we relate the incoming and outgoing scattering modes participating on a single triangle by the $S$ matrices
\begin{align}
\begin{pmatrix}
b_{2}\\
\beta_{2}
\end{pmatrix} &= S_{1} \begin{pmatrix}
a_{1}\\
\alpha_{1}
\end{pmatrix},\label{eq.S1}\\
\begin{pmatrix}
b_{3}\\
\beta_{3}
\end{pmatrix} &= S_{2} \begin{pmatrix}
a_{2}\\
\alpha_{2}
\end{pmatrix},\label{eq.S2}\\
\begin{pmatrix}
b_{1}\\
\beta_{1}
\end{pmatrix} &= S_{3} \begin{pmatrix}
a_{3}\\
\alpha_{3} \label{eq.S3}
\end{pmatrix}.
\end{align}
We have used greek and latin symbols to differentiate between the modes encircling the inner triangle and the outer ones.

Here, the presence of $C_3$ and $\mathcal{M}_x\mathcal{T}$ symmetries imposes that the $S$ matrices used in Eqs.~\eqref{eq.S1}--\eqref{eq.S3} become equal, namely, ${S_{0}:=S_{1}=S_{2}=S_{3}}$ and symmetric $S_{0}=S_{0}^t$, yielding 
the most general (up to a global phase) $2\times2$ unitary symmetric matrix 
\begin{equation}
S_{0}= \begin{pmatrix}
e^{i \varphi}\sqrt{P_R} & i \sqrt{1-P_R}\\
i \sqrt{1-P_R}& e^{-i \varphi}\sqrt{P_R} 
\end{pmatrix}, \label{eq.S0}
\end{equation}
where $P_R=1-P_L$ is the probability to scatter to the right, and $\varphi$ is the phase difference between scattering to the right inside or outside of the triangle depicted in Fig.~\ref{fig.KagomeTriangle}(c). In a realistic situation, we expect $P_R>P_L$ because the geometry of the kagome lattice forces a larger overlap between the incoming and right outgoing wave functions $(60^\circ)$ than for left outgoing wave functions $(120^\circ)$. 
Only in the limit of strongly localized wave functions, the geometry of the lattice stops favoring a right reflection, yielding $P_R \approx P_L\approx0.5$.

The incoming $\alpha_i$ and outgoing $\beta_i$ modes inside the single triangles are related by 
\begin{align}
\alpha_i=	\exp\left[i \pi\epsilon/2  \right] \beta_i, \label{eq.dynphase1}
\end{align}
where $\epsilon = E l/(\pi\hbar v_F)$ is the dynamical phase gathered after propagating along the links, $v_F$ is the Fermi velocity of graphene, and $l/2$ is the length of the link. 

\begin{figure*}
	\includegraphics[width=\textwidth]{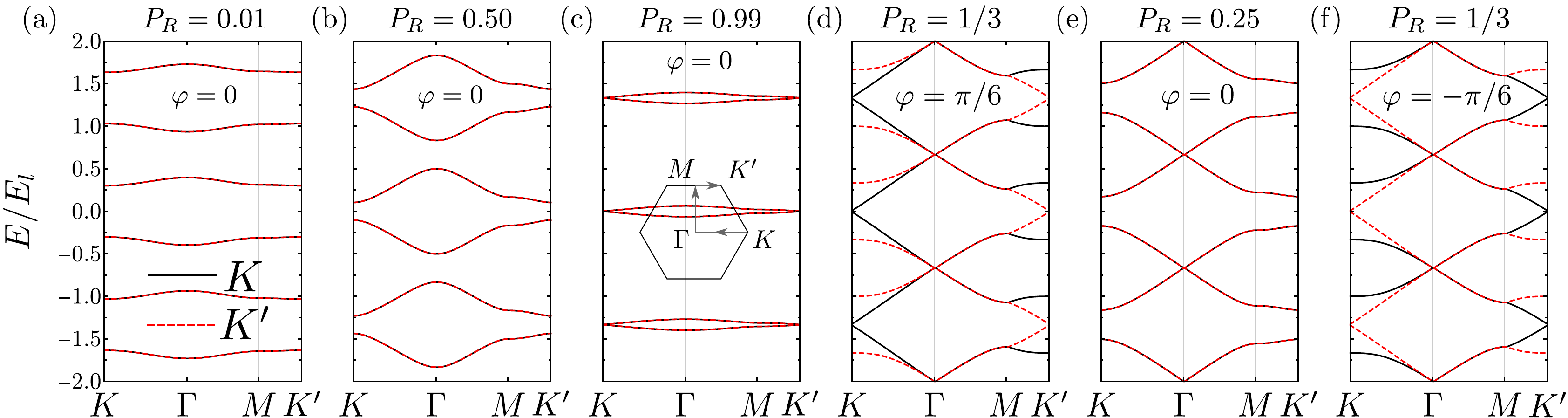}
	\caption{Network bands along high-symmetry lines; see inset in (c) for $K$ (solid black) and $K'$ (red dashed) in units of $E_l=\pi\hbar v_F/l$ for different values of $P_R$ and $\varphi$.} \label{fig.Bands}
\end{figure*}

We simplify the structure of the network by contracting the $S$ matrix of individual triangles, which are encircled in red in Fig.~\ref{fig.KagomeTriangle}(a), onto single scattering nodes. In this way, the kagome chiral network is mapped onto a triangular chiral network [see Fig.~\ref{fig.KagomeTriangle}(b)]. The resulting $S$ matrix of one triangle is energy-dependent due to the dynamical phases picked up along the contracted links. Using Eq.~\eqref{eq.dynphase1}, we combine Eqs.~\eqref{eq.S1}--\eqref{eq.S3} into a single $S$ matrix 
\begin{equation}
\begin{pmatrix}
b_1\\b_2\\b_3
\end{pmatrix} = S_\triangle 	
\begin{pmatrix}
a_1\\a_2\\a_3
\end{pmatrix}
\end{equation}
with 
\begin{widetext}
	\begin{equation}
	S_\triangle = \frac{1}{e^{3i \varphi}-e^{\frac{3 i \pi\epsilon}{2}}P_R^{3/2}}
	\begin{pmatrix}
	-e^{2i \varphi}e^{i \pi\epsilon}P_L \sqrt{P_R} & -e^{3i \varphi}e^{\frac{i\pi\epsilon}{2}}P_L & \left(e^{3i \varphi}-e^{\frac{3i\pi\epsilon}{2}}\sqrt{P_R}\right)e^{i \varphi}\sqrt{P_R}\\
	\left(e^{3i \varphi}-e^{\frac{3i\pi\epsilon}{2}}\sqrt{P_R}\right)e^{i \varphi}\sqrt{P_R} &-e^{2i \varphi}e^{i \pi\epsilon}P_L \sqrt{P_R} & -e^{3i \varphi}e^{\frac{i\pi\epsilon}{2}}P_L\\
	-e^{3i \varphi}e^{\frac{i\pi\epsilon}{2}}P_L
	& \left(e^{3i \varphi}-e^{\frac{3i\pi\epsilon}{2}}\sqrt{P_R}\right)e^{i \varphi}\sqrt{P_R} & -e^{2i \varphi}e^{i \pi\epsilon}P_L \sqrt{P_R}
	\end{pmatrix}, \label{eq.STriangle}
	\end{equation}
\end{widetext}
being the total $S$ matrix of a single triangle. The $S$ matrix for the opposite valley can be obtained using symmetry arguments, that is, $S_{\triangle}^K=(S_{\triangle}^{K'})^t$.

\section{Kagome network without spin-orbit coupling}\label{Sec.wosoi}

In this section, we analyze the energy spectrum and the magnetoconductance of the kagome network without spin-dependent scattering effects. Therefore, the $S$ matrices for spin-up and  -down electrons are identical  and remain decoupled.

\subsection{Network bands}\label{Sec.Bands}

Once we have contracted a single triangle into a scattering center, the kagome network is turned into a triangular network, similar to the one arising in minimally twisted bilayer graphene under an interlayer bias \cite{Prada2013,Efimkin2018, DeBeule2020,DeBeule2021}. 

Thus, to set the triangular network, we place the scattering centers at positions $m \bm{l}_1+n \bm{l}_2$ with $m,n\in \mathbb{Z}$ and $\bm{l}_{1/2}=l/2 (-1/2,\pm\sqrt{3}/2)$.

To obtain the network energy spectrum, we make use of Bloch's theorem, which relates the scattering amplitudes $b_{mn}$ at the node $(m,n)$ to the outgoing scattering amplitudes at different unit cells by
\begin{equation} \label{eq.bloch}
\begin{pmatrix}  b_{1m+1n} \\b_{2m-1n-1}\\  b_{3mn+1} \end{pmatrix}_{\bm k} =  \mathcal M(\bm k)  \begin{pmatrix} b_{1mn} \\ b_{2mn}  \\ b_{3mn}  \end{pmatrix}_{\bm k},
\end{equation}
with $\mathcal M(\bm k) = \textrm{diag} \left( e^{ik_1}, e^{ik_3}, e^{ik_2} \right)$ where $k_j = \bm k \cdot \bm l_j$ ($j=1,2,3$) and $\bm l_3 = -(\bm l_1+\bm l_2)$.
The incoming and outgoing modes of different nodes are related by \cite{Efimkin2018, Pal2019}
\begin{align}
&(a_{1mn},a_{2mn},a_{3mn})^t \label{eq.dynphase}=e^{i\pi\epsilon/2  }(b_{1m+1n},b_{2m-1n-1},b_{3mn+1})^t.
\end{align}
Finally, the energy bands are obtained substituting $b_{mn}=S_\triangle a_{mn}$ and Eq.~\eqref{eq.dynphase} into \eqref{eq.bloch}, leading to
\begin{align} \label{eq.bands}
\mathcal M(\bm k) S_\triangle \, a_{\bm k} = e^{-i\pi\epsilon/2} a_{\bm k}, 
\end{align}
from which we obtain the equation
\begin{equation}
\text{det}\left(\mathcal{M}(k)S_\triangle-e^{-i\pi\epsilon/2} \mathds{1}_3\right) = 0,\label{eq:energieswoSO}
\end{equation}
whose solution gives the network energy bands. Note that $S_\triangle$ is energy-dependent, i.e.,~$S_\triangle=S_\triangle(\epsilon)$.
This equation results in 
\begin{equation}
f(k_x,k_y)=\cos(\frac{3\pi\epsilon}{2})
\end{equation} 
with
\begin{align}
f(k_x,k_y)&=\sqrt{P_R}\Bigg[P_R\cos(3\varphi) -(1-P_R) \cos(k_x l/2-\varphi)\nonumber\\
&-2(1-P_R)\cos(\frac{k_x}{2}\frac{l}{2}+\varphi)\cos(\frac{\sqrt{3}k_y}{2}\frac{l}{2})\Bigg]\label{eq:funf}
\end{align}
with a periodicity of $4/3$ in $\epsilon$.
We find two solutions of this equation within the interval $\epsilon\in [-2/3,2/3]$, namely,
\begin{equation}
\epsilon_\pm=\frac{E_\pm l}{\pi\hbar v_F}=\pm \frac{2}{3\pi} \arccos[f(k_x,k_y)].\label{eq.solbands}
\end{equation}
Here, the periodicity of the network bands is reduced due to the presence of a phase-rotation symmetry \cite{DeBeule22Kagome}.
The smallest energy window that contains all unique solutions is called the fundamental domain \cite{Delplace2017}, which is in our case $\epsilon\in [-2/3,2/3]$. For the parameter $\varphi$ all unique cases are realized in the interval $[-\pi/3,\pi/3]$. This can be seen by noticing that shifting $\varphi$ by $2\pi/3$ in Eq.~\eqref{eq:funf} gives the same result as if we would shift the center of the Brillouin zone $\Gamma$ to $K (K')$, which is consistent with Ref.~\onlinecite{DeBeule22Kagome}. The energy spectrum of the other valley is obtained by $(k_x,k_y)\rightarrow(-k_x,-k_y)$ and $S_{K'}=S^t_\Delta$ [see Eq.~\eqref{eq.STriangle}].
Since Eq.~\eqref{eq:funf} is invariant under $(k_x,k_y)\rightarrow(-k_x,-k_y)$ for $\varphi=0$, the bands for $K$ and $K'$ coincide [see Figs.~\ref{fig.Bands}(a)--\ref{fig.Bands}(c)].

In Fig.~\ref{fig.Bands}, we show the resulting network bands along high-symmetry lines, depicted in the upper right inset in Fig.~\ref{fig.Bands}(c). We start with the case of $\varphi=0$. In the limit $P_R\rightarrow 1(0)$, the combined $S$ matrix $S_\triangle$ has no forward scattering probability and, thus, electrons encircle individual triangles (hexagons), which result in the appearance of flat bands [see Figs.~\ref{fig.Bands}(a) and \ref{fig.Bands}(c)].

For intermediate values of $0<P_R<1$, the band structure acquires a finite group velocity since now there is a finite forward scattering probability that allows a coupling between modes encircling  triangles and hexagons [see Fig.~\ref{fig.Bands}(b)]. However, the spectrum remains gapped in these cases. As a next step, we therefore study in which parameter regimes the gaps become closed. The band structure exhibits a gap closing at the $\Gamma$ point for $P_R=1/4$ [see Fig.~\ref{fig.Bands}(e)]. Indeed, this result can be generalized for $\varphi\neq 0$, where now the gap closing occurs for $P_R= 1/[4 \cos^2(\varphi)]$. In addition, we can find a gap closing at the $K (K')$ point for $P_R= 1/[4 \cos^2(\varphi\pm 2\pi/3)]$ [see Figs.~\ref{fig.Bands}(d)--\ref{fig.Bands}(f)]. The gaps at the $\Gamma$ and $K (K')$ points can get simultaneously closed for $P_R =1/3$ and $\varphi=\pm\pi/6$. 
 We notice, that for finite $\varphi$, the symmetry $k_x\rightarrow-k_x$ becomes lifted and, therefore, the bands of the two valleys are no longer the same for all $\mathbf{k}$. However, along $\Gamma \rightarrow M$ this symmetry is never lifted because the bands remain symmetric under  $k_y\rightarrow-k_y$ for all values of $\varphi$. Furthermore, $\varphi\rightarrow-\varphi$ has the same effect on the bands as $K\rightarrow K'$.

\subsection{Magnetoconductance} \label{Sec.Cond}
We calculate the conductance of a network strip with width $W$ and length $L\ll W$. To this aim, we combine recursively the $S$ matrices along $L$ and sum over the good quantum number $k_y$ in the transversal direction (see further details in Refs.~\onlinecite{DeBeule2020,DeBeule2021, Wittig2023} and Appendix \ref{App.Combining}). 

The magnetoconductance for finite temperature is given by \cite{datta_1995}
\begin{equation}
G = G_0  \frac{W}{\sqrt{3} l/2}\int \dd{E}\mathcal{T}(E) \left(-\frac{\partial f_0}{\partial E}\right),\label{eq.Cond}
\end{equation}
where $G_0=e^2/h$, $W$ is the width of the strip, $f_0=\{\exp[(E-E_F)/k_BT]+1\}^{-1}$ is the Fermi-Dirac distribution with $E_F$ the Fermi energy, and $\mathcal{T}(E)$ is the transmission function for one unit cell of the strip.  At zero temperature, Eq.~\eqref{eq.Cond} reduces to
\begin{equation}
G = G_0  \frac{W}{\sqrt{3} l/2}\mathcal{T}(E_F).
\end{equation}
We only discuss the transmission from the left to the right terminal here, which is the same as from right to left in each valley in these kind of strip systems for $W\gg L$ \cite{DeBeule2020}.

\begin{figure}
	\includegraphics[width=0.48\textwidth]{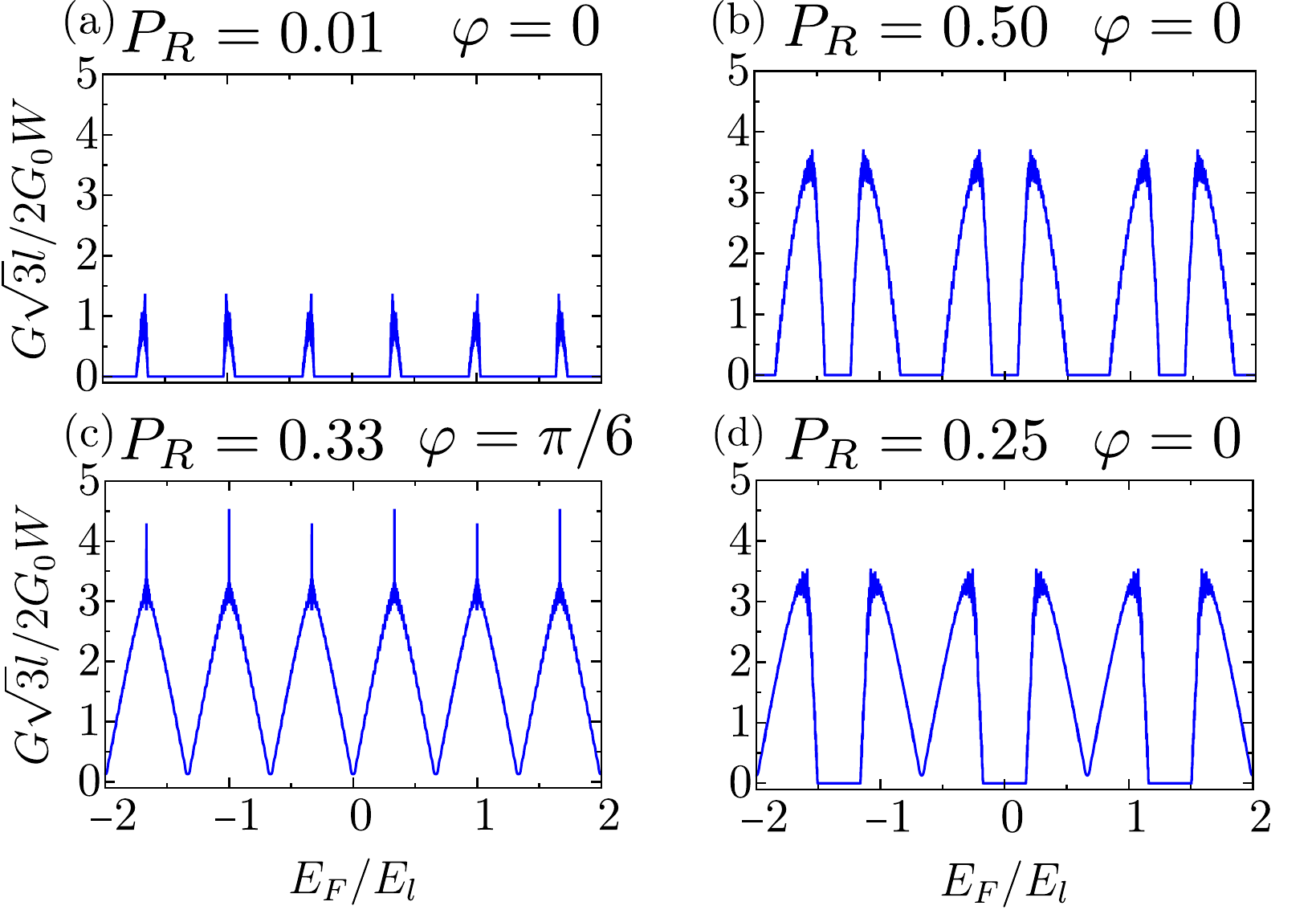}
	\caption{Conductance as a function of $E_F$ of a network strip for different values of $P_R$ and $\varphi$ at zero temperature with $L=10 l$.} \label{fig.Condalpha0}
\end{figure}

{\bf Conductance at $\mathbf{B=0}$}---In Fig.~\ref{fig.Condalpha0}, we present the conductance as a function of the Fermi energy, in units of $E_l=\pi \hbar v_F/l$. As expected, the conductance reaches its maximum at $P_R = 0.5$ [Fig.~\ref{fig.Condalpha0}(b)], as both right and left scattering contributions are needed for transport through the network. We observe an energy periodicity of $4/3E_l$, similarly to the network bands (see Fig.~\ref{fig.Bands}). Additionally, the conductance vanishes at $P_R\rightarrow0$ and $P_R\rightarrow1$ [Fig.~\ref{fig.Condalpha0}(a)], corresponding to the appearance of flat bands discussed in Figs.~\ref{fig.Bands}(a) and \ref{fig.Bands}(c). 
We show in Figs.~\ref{fig.Condalpha0}(c) and \ref{fig.Condalpha0}(d) two examples of the conductance that exhibit gap closings in the corresponding bands Figs.~\ref{fig.Bands}(d) and \ref{fig.Bands}(e). In the case of $P_R=1/3, \varphi=\pi/6$ [Fig.~\ref{fig.Condalpha0}(c)], we observe a reduction of the periodicity of the conductance to $2/3 E_l$, which is half of the previous periodicity. Interestingly, we observe conductance peaks at $E_F = \pm (2n/3+1/3)E_l$ with $n\in \mathds{N}$, which are higher than the peaks observed for $P_R=0.5$ and $\varphi=0$. This resonance phenomenon occurs for $\varphi=\pm(2n+1)\pi/6$ with $n\in \mathds{N}$.

To have more insight into this resonance phenomenon, we construct a reduced version of the network model consisting of three contracted triangles [see Fig.~\ref{fig.KagomeTriangle}(b)]. To simplify the calculations we consider periodic boundary conditions in the vertical direction of the strip. Under these conditions, the phase factors in the transmission function are all of the form $\exp(3im_1 \varphi)\exp(3im_2\pi\epsilon_F/2)$ with $\epsilon_F=E_F/E_l$, $m_1,m_2\in\mathds{N}$, and $m_1+m_2\in 2\mathds{N}$. For $\varphi=\pm(2n_1+1)\pi/6$ and $\epsilon_F=\pm (2n_2/3+1/3)$, with $n_1,n_2\in \mathds{N}$, 
these phase factors become all $\pm 1$, indicating an interference effect. The transmission function for these values of $\epsilon_F$ and $\varphi$ is for a single valley and spin
\begin{align}
    \mathcal{T}=2+\frac{25}{\left(5\pm\sqrt{P_R}+P_R\right)^2}-\frac{10}{5\pm\sqrt{P_R}+P_R}. \label{eq.minimal}
\end{align}
We observe a constant term in the transmission function, independent of $P_R$, that lead to the discussed peaks in Fig.~\ref{fig.Condalpha0}.

{\bf Conductance at $\mathbf{B\neq0}$}---
We introduce the effects of a perpendicular magnetic field by means of the shift of momentum by a vector potential $\mathbf{A}=Bx \mathbf{e}_y$. As a result, the modes acquire a Peierls phase, which is proportional to the magnetic flux $\Phi=B \mathcal{A}$. Here, $\mathcal{A}=\sqrt{3}l^2/2$ represents the area of the structural element of the kagome network, consisting of a hexagon ($\mathcal{A}_{\hexagon}=3\sqrt{3}l^2/8$) and two triangles ($\mathcal{A}_\triangle=\sqrt{3}l^2/16$) [see Fig.~\ref{fig.KagomeTriangle}(a)]. It is important to note that the $S$ matrix for one triangle is modified in the presence of a magnetic field. We provide its analytical expression in Appendix \ref{App.triangularalpha0} in Eq.~\eqref{eq.Flux}.

\begin{figure}
	\includegraphics[width=0.48\textwidth]{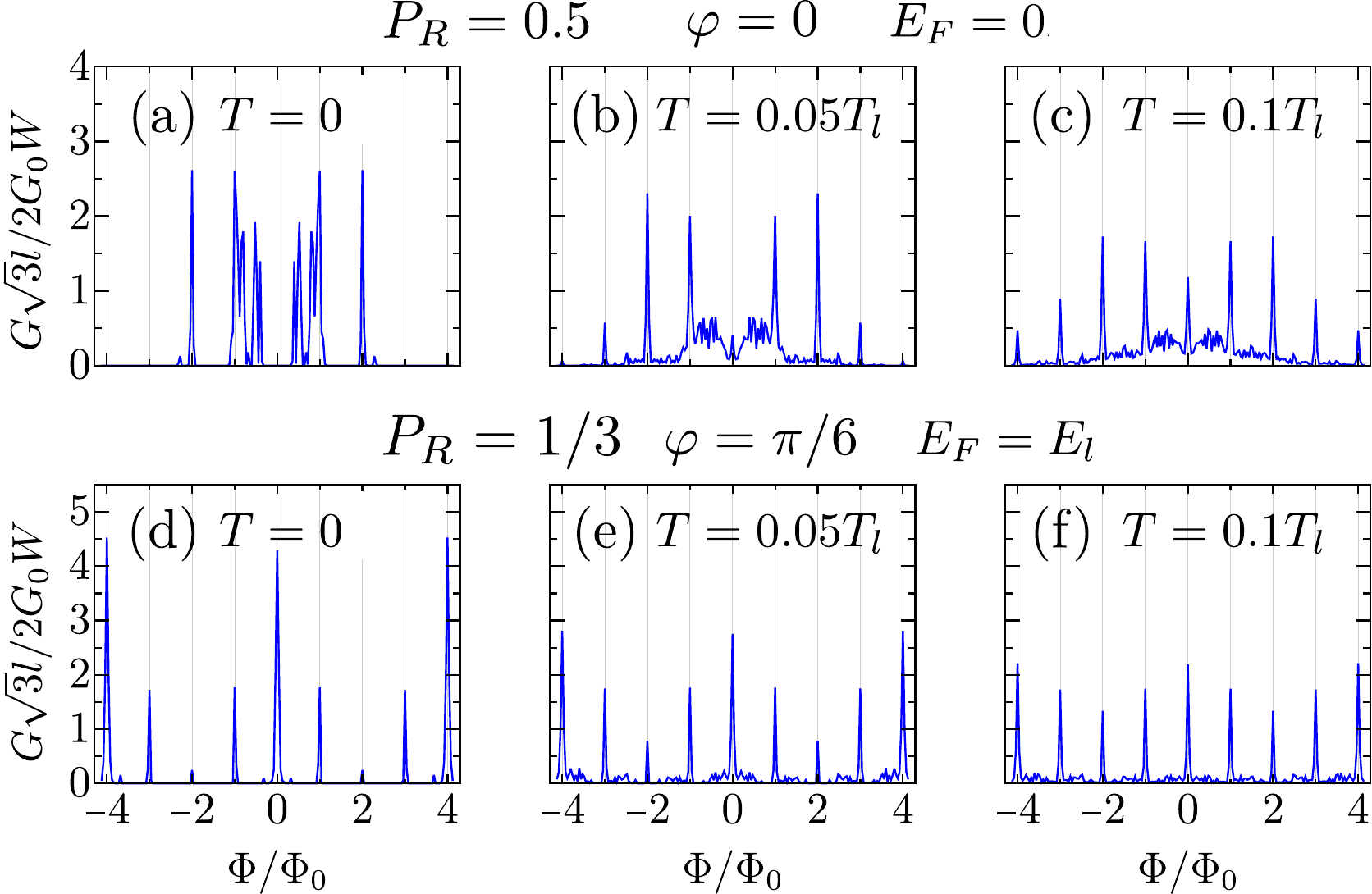}
	\caption{Conductance as a function of $\Phi$ of a network strip for $P_R=0.5$, $\varphi=0$, $E_F=0$ (top) and $P_R=1/3$, $\varphi=\pi/6$, $E_F=E_l$ (bottom) for different temperatures and with $L=5l$. Here, $E_l=\pi \hbar v_F/l$ and $T_l=\pi\hbar v_F/(k_B l)$.} \label{fig.MagnetoNoSO}
\end{figure}

\begin{figure}
	\includegraphics[width=0.48\textwidth]{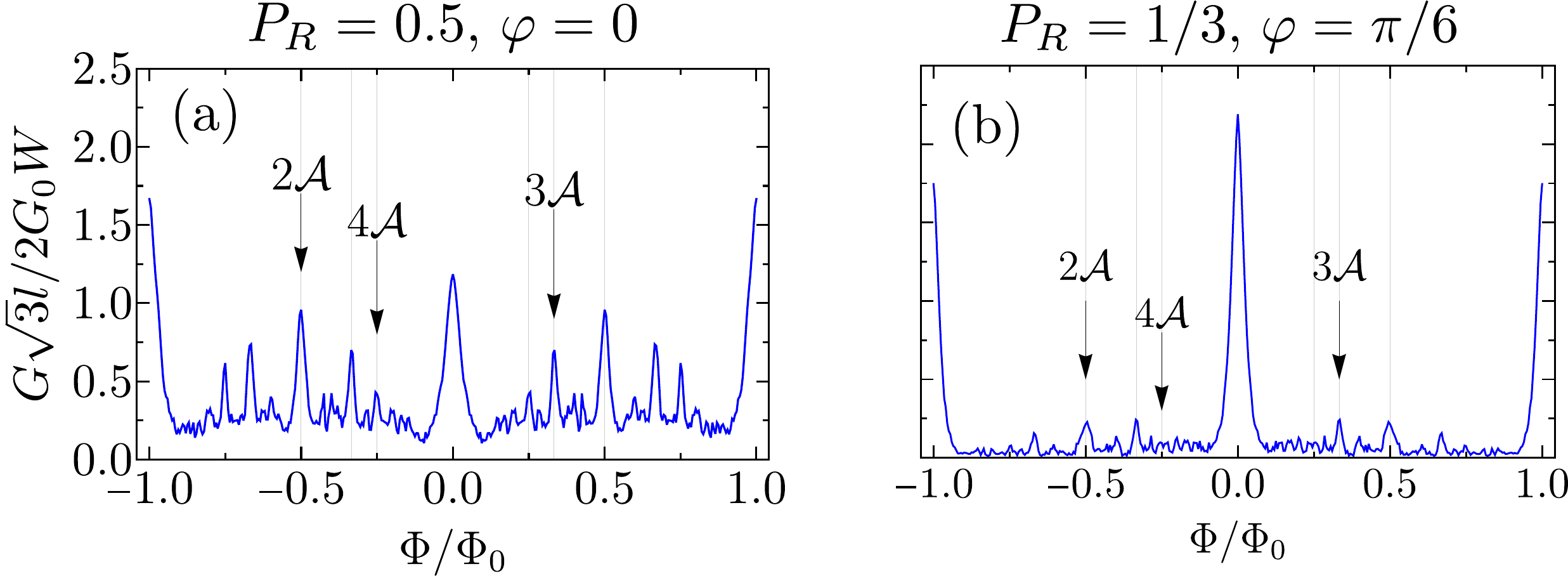}
	\caption{Conductance as a function of $\Phi$ of a network strip for $P_R=0.5$, $\varphi=0$, $E_F=0$ (left) and $P_R=1/3$, $\varphi=\pi/6$, $E_F=E_l$ (right) for $T/T_l=0.1$ and with $L=5l$. Here, $E_l=\pi \hbar v_F/l$ and $T_l=\pi\hbar v_F/(k_B l)$.} \label{fig.MagnetoNoSOZoom}
\end{figure}

We show the magnetoconductance in Fig.~\ref{fig.MagnetoNoSO}, for different values of $P_R$, $\varphi$ and temperature $T$. For finite $\Phi\neq0$, the conductance has a periodicity of $\Phi/\Phi_0=8$, which arises due to the possibility of encircling a single triangle with an area $\mathcal{A}_\triangle$ that is an eighth of the area of the unit cell of the kagome network $\mathcal{A}$.

Electrons propagating through the network gather dynamical and Peierls phases, which can lead to destructive and constructive interference between different trajectories. Aharonov-Bohm (AB) conductance resonances occur when electronic paths encircle multiples of the unit cell of the kagome network, with an area $\mathcal{A}$. Thus, these paths  have all the same length and, therefore, they accumulate the same dynamical phase, yielding an energy-independent transmission probability. This has a crucial consequence at finite temperature since paths with different lengths accumulate different dynamical phases are averaged out in energy \cite{Virtanen2011,DeBeule2020}.  We see this effect in Fig.~\ref{fig.MagnetoNoSO}, where the  AB conductance resonances become more prominent at finite temperature for $\Phi=n\Phi_0$ with $n\in\mathds{Z}$. Aside from the AB resonances, we also observe an energy-dependent background, which becomes less visible with higher temperature and/or larger system size and depends on the value of the Fermi energy $E_F$.

In addition, it is also possible to encircle larger areas in this network, which would result in additional peaks in the magnetoconductance at fractions of $\Phi_0$. More specifically, encircling an area of $n \mathcal{A}$ with $n\in \mathds{N}$ would lead to a peak at $\Phi/\Phi_0=1/n$. To make this effect visible,  we show in Fig.~\ref{fig.MagnetoNoSOZoom} the results from Fig.~\ref{fig.MagnetoNoSO} in a smaller range of flux $\Phi$ for finite temperature $T/T_l=0.1$. We find between two larger peaks at integer multiples of $\Phi_0$ also smaller peaks due to encircling higher multiples of $\mathcal{A}$, as expected. 
Note that these conductance peaks are smaller because it is less probable to encircle larger areas.

\section{Interplay of Rashba spin-orbit coupling and the kagome network}\label{Sec.SO}
Next, we explore the influence of Rashba spin-orbit (SO) coupling on the spectrum and transport properties of the network. To do so we have to include the presence of SO coupling into the network calculations. The two primary examples or proposed kagome networks are based on graphene systems, namely, an hBN-graphene-hBN heterostructure with a relative twist angle between each layer \cite{Vladimir2022Kagome} and a single graphene layer in a periodic strain field \cite{DeBeule22Kagome}. Therefore, we model the system with the graphene Hamiltonian. For low energies the Hamiltonian around the $K/K'$ valley with Rashba spin-orbit coupling \cite{Kane2005, Hongki2006, Kochan2017} is given by
\begin{align}
    H&= H_0 + H_R,\\
    H_0&= \hbar v_F \left(\tau k_x \sigma_x+k_y \sigma_y\right),\\
    H_R&= \alpha_R (\tau \sigma_x s_y - \sigma_y s_x),
\end{align}
with $s_i$ ($\sigma_i$) the $i$th Pauli matrix in spin and sublattice spaces,  $\alpha_R$ the SO coupling constant, and $v_F$ the Fermi velocity of graphene.
The factor $\tau=\pm1$ takes different values for the $K/K'$ valley. We can rewrite this Hamiltonian as a spin-dependent shift ($\bm{s}\cross \bm{\alpha}_R $) in momentum, namely,
\begin{align}
    H= \hbar v_F \bm{\sigma} \cdot \left( \bm{k}+\bm{s}\cross \bm{\alpha}_R\right) \label{eq.HSO}
\end{align}
with $\bm{\alpha}_R=\big(0,0,\alpha_R/(\hbar v_F)\big)^t$, $\bm s=(s_x,s_y,s_z)^t$, $\bm{\sigma}= (\tau\sigma_x,\sigma_y,\sigma_z)^t$, and $\bm{k}=(k_x,k_y,0)^t$. Note that we have changed the notation here to an equivalent representation with three-dimensional vectors. 
Similar to the effect of a magnetic field \cite{Aharonov1959,Bruus2004} that leads to the Peierls phase, we can describe this momentum shift as a geometric phase picked up along the links in the form of  
\begin{align}
\tilde{\psi}(\bm{r})&=  \exp(-i \int_{C} \left(\bm{s}\cross \bm{\alpha}_R\right)  \dd{\bm{l}})\psi(\bm{r})\\
&=  \exp(-i \left(\bm{s}\cross \bm{\alpha}_R\right) \bm{r})\psi(\bm{r})=: A(\theta) \psi(\bm{r}), \label{eq.geophase}
\end{align}
where $\bm{r}=(x,y,z)^t$ describes the end point of the curve,  $\psi$ is the eigenfunction of Eq.~\eqref{eq.HSO} for $\alpha_R=0$ and $\tilde{\psi}$ for finite spin-orbit coupling. We assumed here that the start point of the curve $C$ is zero. We can write the geometric phase factor in Eq.~\eqref{eq.geophase} as
\begin{align}
A(\theta)&= \exp\left[i \pi \alpha (s_x\sin(\theta)-s_y\cos(\theta))/2\right] \label{eq.CasherPhase}
\end{align}
with $\alpha=\alpha_R l/(\pi \hbar v_F)$ and $\theta$ the angle between the direction of the mode relative to the $x$ axis (see Fig.~\ref{fig.KagomeTriangle}). Note that this form is similar to the way it was implemented in previous publications in different network systems \cite{Bercoiux2004,Bercioux2005}. To give an estimate for the value of $\alpha$ we find in the literature that for graphene on $\text{WSe}_2$ a Rashba spin-orbit coupling constant of $\alpha_R\approx 15$ meV \cite{Sun2023} has been measured. Furthermore, it is reasonable in a moir\'e system with small twist angle to assume that $l\sim100$ nm \cite{DeBeule2021, Vladimir2022Kagome, Wittig2023}. With the graphene Fermi velocity $v_F\approx 10^6$ m/s \cite{DeBeule2021} we find $\alpha \sim 0.7$.

\begin{figure*}
	\includegraphics[width=\textwidth]{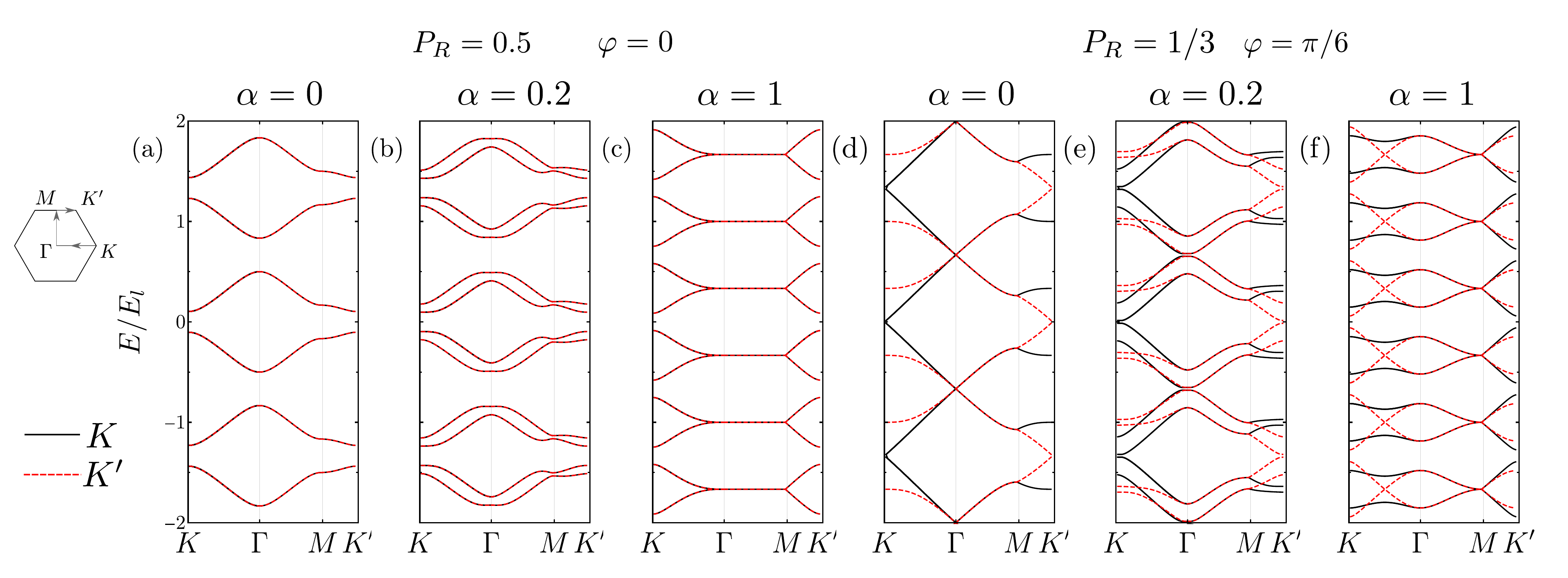}
	\caption{Network bands along high-symmetry lines of the kagome network for $P_R=0.5$ and $\varphi=0$ (a)--(c) and $P_R=1/3$ and $\varphi=\pi/6$ (d)--(f) for different values of spin-orbit coupling $\alpha$.} \label{fig.SOBands}
\end{figure*}

The outgoing modes relate to the incoming modes inside the triangle (see notation in Fig.~\ref{fig.KagomeTriangle}) as
\begin{align}
(\alpha_{1,\uparrow},\alpha_{1,\downarrow})^t&= M(2\pi/3)(\beta_{1,\uparrow},\beta_{1,\downarrow} )^t, \label{eq.dyn1}\\
(\alpha_{2,\uparrow},\alpha_{2,\downarrow})^t&= M(0)(\beta_{2,\uparrow},\beta_{2,\downarrow} )^t, \label{eq.dyn2}\\
(\alpha_{3,\uparrow},\alpha_{3,\downarrow})^t&= M(-2\pi/3)(\beta_{3,\uparrow},\beta_{3,\downarrow} )^t, \label{eq.dyn3}
\end{align}
by means of the matrix 
\begin{align}
    M(\theta)&= \exp\left[ i \pi  s_0\epsilon/2\right]A(\theta) \label{eq.MSO}
\end{align}
 with $s_0$ the identity matrix in spin space.

We have now everything to describe a single triangle of the kagome network in the presence of Rashba spin-orbit coupling. Thus, we replace the dynamical phase introduced in Eq.~\eqref{eq.dynphase} by Eqs.~\eqref{eq.dyn1}--\eqref{eq.dyn3} and calculate the network bands  and magnetoconductance calculations in Secs.~ \ref{Sec.BandsSO} and~\ref{Sec.CondSO}, respectively.

\subsection{Network bands}\label{Sec.BandsSO}

Instead of contracting each triangle into an energy-dependent scattering node, we enlarge our basis set to account for all incoming and outgoing modes at each subnode inside the triangle, yielding a larger, but energy-independent $S$ matrix. As it was pointed out in Ref.~\onlinecite{DeBeule22Kagome}, this way of calculating the energy spectrum makes the calculation of the network bands more stable. 

Thus, using the notation of the incoming and outgoing modes depicted in Fig.~\ref{fig.KagomeTriangle}, we write for a given spin $\sigma$ 
\begin{align}
&(b_{2,\sigma},\beta_{2,\sigma},b_{3,\sigma},\beta_{3,\sigma},b_{1,\sigma},\beta_{1,\sigma})^t\nonumber\\
&=S_\sigma (a_{1,\sigma},\alpha_{1,\sigma},a_{2,\sigma},\alpha_{2,\sigma},a_{3,\sigma},\alpha_{3,\sigma})^t
\end{align} 
with
\begin{equation}
S_\sigma = \begin{pmatrix}
S_{0,\sigma} &0 &0\\
0 & S_{0,\sigma} & 0\\
0 & 0 & S_{0,\sigma}
\end{pmatrix}. \label{eq.ssigma}
\end{equation}
Here, we have assumed that the $S$ matrix at every node is spin-independent, so that Eq.~\eqref{eq.S0} is still valid, i.e., $S_{0,\sigma}=S_0$.

For the next step, we label each of these triangles, which describe the unit cell of the kagome network, by its position $m \bm{l}_1+n \bm{l}_2$ with $m,n\in \mathds{Z}$ and $\bm{l}_{1/2}=l/2 (-1/2,\pm\sqrt{3}/2)$. Then, we can relate the incoming scattering amplitudes at a unit cell $(m,n)$ to the outgoing scattering amplitudes $b_{mn}$ at different unit cells by
\begin{align}
(a_{1mn,\uparrow},a_{1mn,\downarrow})^t&= M(-\pi/3)(b_{1m+1n,\uparrow},b_{1m+1n,\downarrow} )^t, \label{eq.dyn4}\\
(a_{2mn,\uparrow},a_{2mn,\downarrow})^t&=M(\pi)(b_{2m-1n-1,\uparrow},b_{2m-1n-1,\downarrow} )^t, \label{eq.dyn5}\\
(a_{3mn,\uparrow},a_{3mn,\downarrow})^t&= M(\pi/3)(b_{3mn+1,\uparrow},b_{3mn+1,\downarrow} )^t. \label{eq.dyn6}
\end{align}
These equations, together with Eqs.~\eqref{eq.dyn1}--\eqref{eq.dyn3}, allow to write the matrix 
\begin{align}
	D(\alpha,\epsilon)= \text{diag} [ M(-\pi/3),  &M(2\pi/3), M(\pi),M(0),\nonumber \\
 &M(\pi/3),M(-2\pi/3)]
  \label{eq.D}
	\end{align}
containing the dynamical and geometric phases \cite{AharonovCasher1984,Bercioux2015} gathered when propagating between the nodes specified in the corresponding equations (see more details in Appendix \ref{App.Bands}).

Analogously to Eq.~\eqref{eq.bands}, we use Bloch's theorem to obtain
\begin{align} 
\mathcal{M}(\bm{k}) S \, a_{\bm{k}} = D(\alpha,\epsilon)^{-1} a_{\bm{k}}. \label{eq.Bands1}
\end{align}
In contrast to Eq.~\eqref{eq.bands}, here, the $S$ matrix accounts for both spins $S=S_\sigma\otimes\mathds{1}_{2\cross2}$ and the dynamical phase is replaced by the matrix $D(\alpha,\epsilon)$, introduced previously. Moreover, the specific form of the matrix $\mathcal M(\bm k)$ is given in Appendix \ref{App.Bands}.

We obtain the $S$ matrix of the other valley by performing a time-reversal transformation, that is, $M(\theta)\rightarrow M(\theta+\pi)$ and $(k_x,k_y)\rightarrow (-k_x,-k_y)$. In addition, $S$ remains invariant because it is symmetric $S=S^t$.

The energy bands for a given valley are obtained from the solutions of
\begin{align}
\det[\mathcal M(\bm k) S- D(\alpha,\epsilon)^{-1}] =0.\label{eq.Bands2}
\end{align} 

In Fig.~\ref{fig.SOBands} we show the energy bands along high-symmetry lines as depicted in the upper left inset. We can observe the typical spin-momentum splitting in the bands together with a band flattening, which indicates a localization effect (see also 
Refs.~\onlinecite{Bercoiux2004,Bercioux2005}). Additionally, we find for $\alpha=1$ and $\varphi=0$, bands, that are constant from $\Gamma\rightarrow M$, see Fig.~\ref{fig.SOBands}(c). These flattened bands will lead to conductance resonances, which we discuss next. 

\subsection{Magnetoconductance} \label{Sec.CondSO}
We calculate the conductance of the spinful network by combining recursively  $S$ matrices of consecutive single triangles, as specified in Appendix~\ref{App.Combining}. Again, the conductance is obtained with Eq.~\eqref{eq.Cond}. Due to the basis enlargement and the presence of off-diagonal terms introduced by Eqs.~\eqref{eq.ssigma} and \eqref{eq.dyn1}--\eqref{eq.dyn3}, we are forced to calculate $S_\triangle$ numerically, which relates
\begin{equation}
\begin{pmatrix}
b_{1,\uparrow}\\b_{1,\downarrow}\\b_{2,\uparrow}\\b_{2,\downarrow}\\b_{3,\uparrow}\\b_{3,\downarrow}
\end{pmatrix} = S_\triangle 	
\begin{pmatrix}
a_{1,\uparrow}\\a_{1,\downarrow}\\a_{2,\uparrow}\\a_{2,\downarrow}\\a_{3,\uparrow}\\a_{3,\downarrow}
\end{pmatrix}. \label{eq.SmatrixSO2}
\end{equation}
We obtain the $S$ matrix for the other valley performing a spinful time-reversal symmetry transformation, namely, 
\begin{align}
S_{\triangle,K'}(\Phi)= g S^t_{\triangle,K}(-\Phi) g^{-1}, \label{eq.trsScatt}
\end{align}
where $g=-i (\mathds{1}_3 \otimes \sigma_y)$ and $\Phi$ is the flux.

\begin{figure}
	\includegraphics[width=0.48\textwidth]{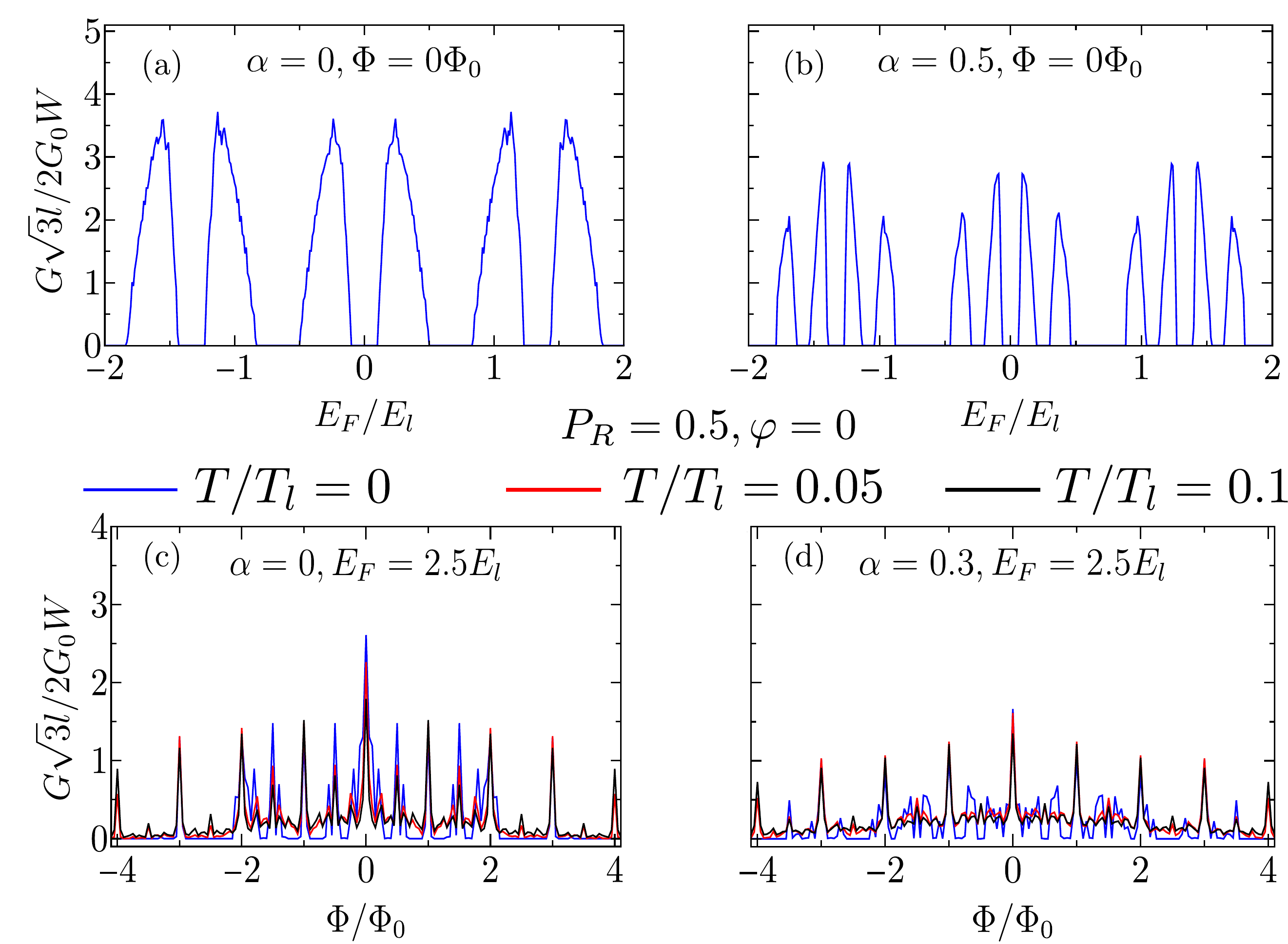}
	\caption{(a), (b) Conductance of a network strip of the kagome network as a function of $E_F$ in units of $E_l=\pi \hbar v_F/l$ with different values of spin-orbit coupling $\alpha$. We show here the case of $P_R=0.5,\varphi=0$ and with $L=10l$. (c), (d) Magnetoconductance of a network strip of the kagome lattice with different values of spin-orbit coupling $\alpha$ as a function of $\Phi$ for different temperatures, $P_R=0.5,\varphi=0$ and length $L=5l$. $E_F$ is given in units of $E_l=\pi \hbar v_F/l$ and temperature $T$ in units of $T_l=\pi\hbar v_F/(k_B l )$.} \label{fig.SOcond}
\end{figure}

In Figs.~\ref{fig.SOcond}(a) and \ref{fig.SOcond}(b), we can observe the impact of the presence of a finite SO coupling on the conductance of the network strip as a function of $E_F/ E_l$ for $P_R=0.5$, $\varphi=0$, $\Phi=0$. We observe a larger number of peaks with a reduced height. These results are consistent with the split bands with a flattened dispersion due to the SO coupling, as seen in Fig.~\ref{fig.SOBands}. 

Next, we study the interplay of the spin-orbit coupling and a perpendicular magnetic field defined by a vector potential of the form $\mathbf{A}=Bx \mathbf{e}_y$ as we have introduced in Sec.~\ref{Sec.Cond}. 
We show the conductance as a function of the magnetic flux $\Phi/\Phi_0$ in Figs.~\ref{fig.SOcond}(c) and \ref{fig.SOcond}(d), with $P_R=0.5$, $\varphi=0$, $E_F=2.5E_l$, and different values of $\alpha_R$ and temperature. For $\alpha=0$, we observe a periodicity of $\Phi=8\Phi_0$ and the AB resonances occurring at higher temperatures at $\Phi=m\Phi_0$ with $m\in \mathds{Z}$ [see Fig.~\ref{fig.SOcond}(c)]. For higher values of $\alpha$ [Fig.~\ref{fig.SOcond}(d)], the AB resonances are diminished, indicating a localization effect of the SO coupling. This reduction makes it challenging to differentiate the resonances at multiples of $\Phi_0$ from the higher-order resonances in-between.

\begin{figure}
	\includegraphics[width=0.48\textwidth]{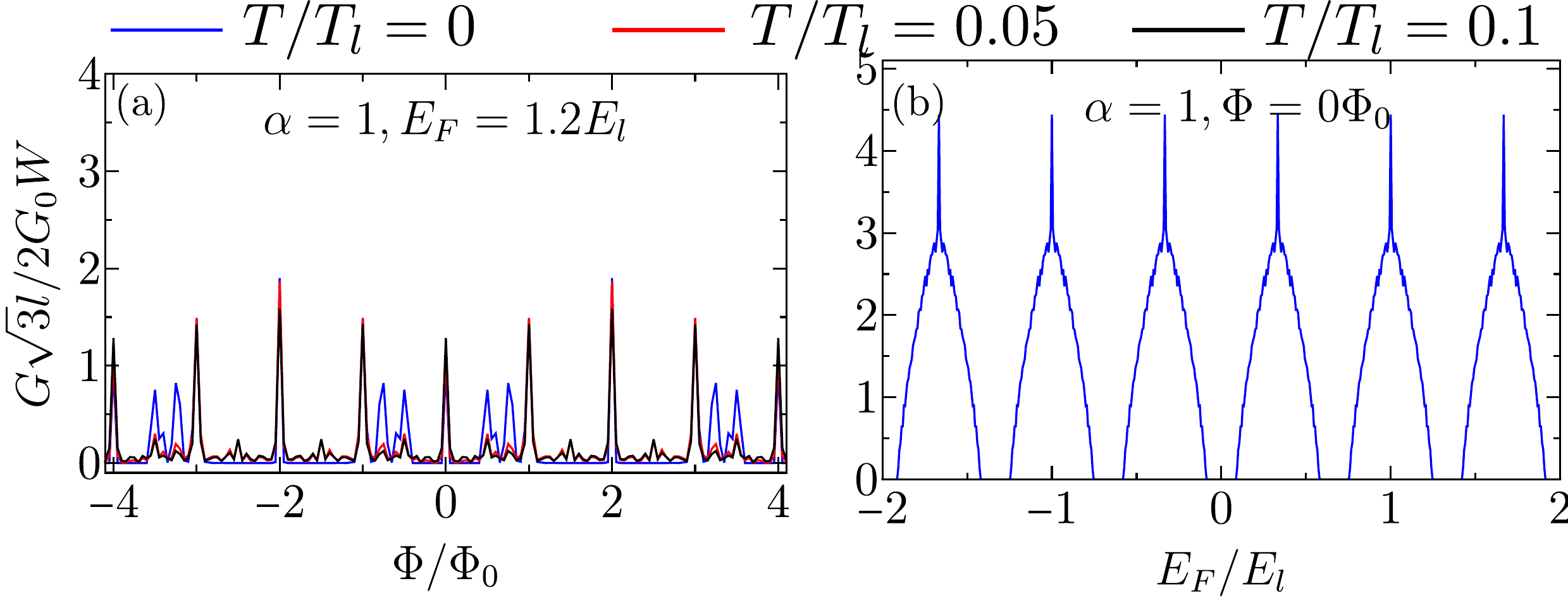}
	\caption{(a) Conductance as a function of $\Phi$ of a network strip of the kagome network with $\alpha=1$ for different temperatures. $E_F$ is given in units of $E_l=\pi \hbar v_F/l$ and temperature $T$ in units of $T_l=\pi\hbar v_F/(k_B l )$. (b) Conductance  of a network strip of the kagome network as a function of $E_F$ in units of $E_l=\pi \hbar v_F/l$ for $\alpha=1$. We show here the case of $P_R=0.5,\varphi=0$ and with $L=10l$. } \label{fig.Res}
\end{figure}

\begin{figure*}
	\includegraphics[width=\textwidth]{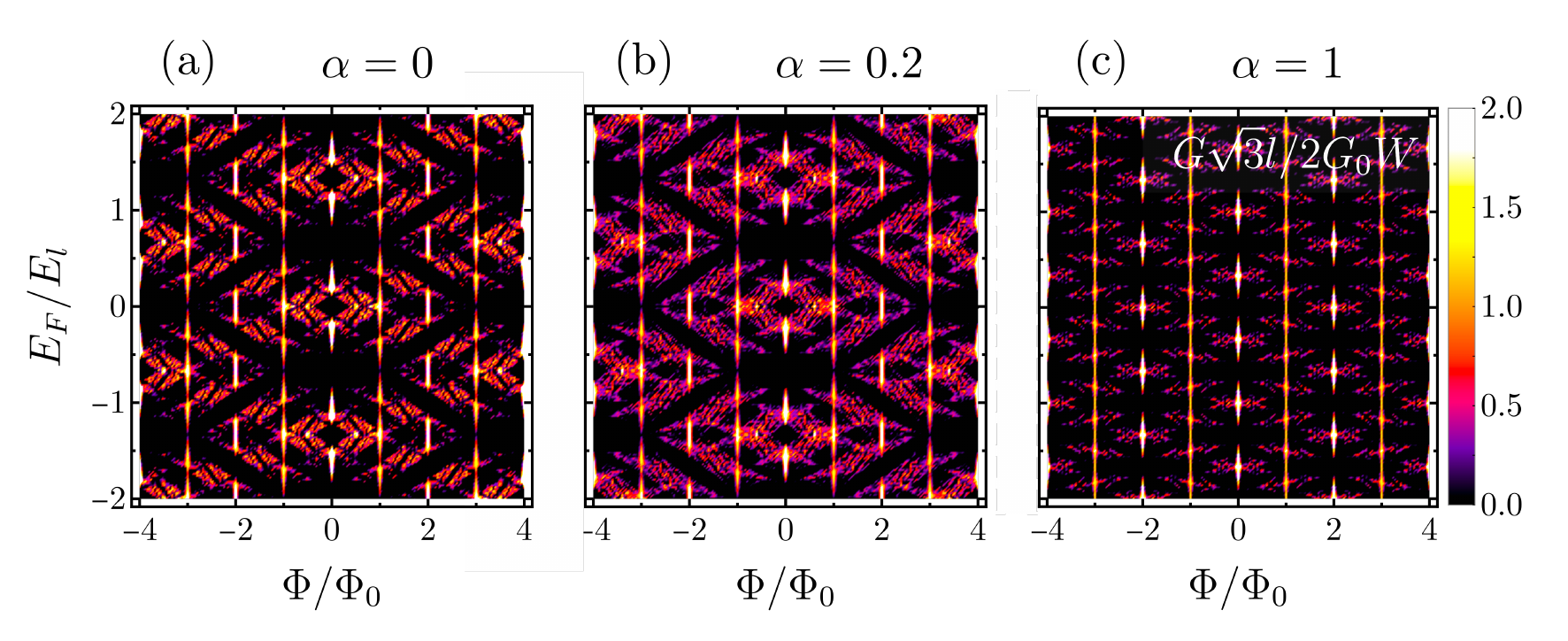}
	\caption{Conductance as a function of $\Phi$ and $E_F$ of a network strip of the kagome network in units of $E_l=\pi \hbar v_F/l$ and flux $\Phi$ in units of $\Phi_0$ with different values of spin-orbit coupling $\alpha$. We show here the case of $P_R=0.5,\varphi=0$ and with $L=5l$.} \label{fig.SODensityPlot}
\end{figure*}

The interplay of the network geometry and the SO is maximally visible when the spin of a particle propagating between two nodes rotates $180^\circ$. Due to the triangular geometry, the effective periodicity of the lattice is doubled since an electron traveling through the network needs to encircle two triangles to return to the same state as before the propagation. This occurs when $M(\theta)$, given in  Eq.~\eqref{eq.MSO}, is completely off-diagonal, which appears for
\begin{align}
   	\alpha = 2n+1,\label{eq.cond1}
\end{align}
with $n\in \mathds{Z}$. We show in Fig.~\ref{fig.Res}(a) the magnetoconductance, where Eq.~$\eqref{eq.cond1}$ is fulfilled. Remarkably, the periodicity of the magnetoconductance is reduced to $\Phi=4\Phi_0$, which indicates that paths encircling a single triangle no longer contribute to the conductance.  Such geometry-dependent interference effects are typical for non-Abelian phase fields, like in this case due to SO coupling, which are responsible for the Aharonov-Casher effect \cite{Bercioux2015}.

Additionally, a resonance phenomenon occurs if apart of fulfilling Eq.~\eqref{eq.cond1}, we also have
\begin{align}
	\frac{3}{2}\epsilon_F \pi - \frac{\pi \Phi}{4 \Phi_0}= (2m+1)\frac{\pi}{2}, \, m \in \mathds{Z}. \label{eq.cond2}
\end{align}
where $\epsilon_F=E_F/E_l$.
When both conditions are fulfilled and $\Phi/\Phi_0$ is an integer, a sharp conductance peak arises [see Fig.~\ref{fig.Res}(b)]. 

We can understand this resonance phenomenon by calculating analytically the probability amplitude of a particle crossing a small version of the network, i.e., a single triangle. The propagator for a single round trip around a triangle is given by $P_1=\exp(-i \pi \Phi/4\Phi_0)M(2\pi/3)M(-2\pi/3)M(0)$ for a contracted triangle or $P_2=\exp(-i \pi\Phi/4\Phi_0)M(-\pi/3)M(\pi/3)M(\pi)$ for an outer triangle, see Fig.~\ref{fig.KagomeTriangle}. Thus, summing over paths with different number of round trips leads to
\begin{equation}
    \sum_{n=0}^\infty P_j^n = \left(\mathbb{1}-P_j\right)^{-1},\, j=1,2.
\end{equation}
This expression exhibits divergences if both conditions are fulfilled, indicating a resonance effect. For finite temperature, the periodicity remains the same [see Fig.~\ref{fig.Res}(a)]. Similar as in Eq.~\eqref{eq.minimal}, we can calculate the transmission function for a given valley for a small system consisting of three triangles [see Fig.~\ref{fig.KagomeTriangle}(b)] with periodic boundary conditions. If Eqs.~\eqref{eq.cond1} and \eqref{eq.cond2} are fulfilled, we find
\begin{align}
    \mathcal{T}=\frac{1250+2P_R\{475+P_R[147+2P_R(13+P_R)]\}}{[25+P_R(9+P_R)]^2}.
\end{align}
We notice again, like in the case of Eq.~\eqref{eq.minimal}, that we have a term in the transmission function that remains finite in the limit of $P_R\rightarrow0(1)$, leading to the observed resonances.

To summarize our findings, we represent the conductance of the network strip as a function of the Fermi energy $E_F$ and magnetic flux $\Phi$ for different values of the SO coupling strength $\alpha$ in Fig.~\ref{fig.SODensityPlot}. We observe that the Hofstadter pattern  \cite{Hofstadter1976,DeBeule2020,DeBeule2021} becomes distorted due to the finite SO coupling. More concretely, we observe a progressive reduction of certain areas of the magnetoconductance by the increase of $\alpha$ [see Figs. \ref{fig.SODensityPlot}(a)--\ref{fig.SODensityPlot}(c)].
These results show the localization effect of the SO coupling observed in other systems \cite{Bercioux2005}.

\subsection{Spin polarization} \label{Sec.Pol}
We now study the spin-dependent transmission, which can have applications in the field of spintronics. The $S$ matrix of the full network can be written in the form 
\begin{equation}
\begin{pmatrix}
b_{1,\uparrow}\\b_{1,\downarrow}\\b_{2,\uparrow}\\b_{2,\downarrow}\\b_{3,\uparrow}\\b_{3,\downarrow}\\b_{4,\uparrow}\\b_{4,\downarrow}
\end{pmatrix} = S 	
\begin{pmatrix}
a_{1,\uparrow}\\a_{1,\downarrow}\\a_{2,\uparrow}\\a_{2,\downarrow}\\a_{3,\uparrow}\\a_{3,\downarrow}\\a_{4,\uparrow}\\a_{4,\downarrow}
\end{pmatrix} \label{eq.SmatrixSO2b}
\end{equation}
(see Fig.~\ref{fig.network}). Similarly as in Eq.~\eqref{eq.trsScatt}, time-reversal symmetry relates
\begin{align}
S_{K'}(\Phi)= g S^t_{K}(-\Phi) g^{-1}, \label{eq.trsScatt2}
\end{align}
where $g=-i (\mathds{1}_4 \otimes \sigma_y)$ and $\Phi$ is the flux. With this matrix we can calculate the transmission function of the network. To analyze the spin polarization, we split the transmission function into its spin components $\mathcal{T}_{nm}=\sum_{j=K,K'}\sum_{\sigma,\sigma'}\mathcal{T}^j_{n\sigma,m \sigma'}$, which obey
\begin{align}
	\mathcal{T}^K_{n \sigma,m \sigma'}(\Phi)= \mathcal{T}^{K'}_{m\sigma',n\sigma}(-\Phi), \label{eq.spintrans}
\end{align}
with $n,m=R,L$ for the right and left leads and $\sigma,\sigma'=\uparrow(\widehat{=}1),\downarrow(\widehat{=}-1)$, respectively. This relation is a direct consequence of Eq.~\eqref{eq.trsScatt2}, which can be written component wise in the form 
\begin{equation}
s^{K}_{i\sigma, j \sigma'}(\Phi)=\sigma \sigma' s^{K'}_{j\sigma', i \sigma}(-\Phi),
\end{equation} 
with $i,j =1,2,3,4$ denote the modes (see Fig. \ref{fig.network}). A similar expression was obtained in a different context of time-reversal symmetric scattering without valley degree of freedom \cite{Zhai2005, Jacquod2012}. Also, we notice that the full transmission function fulfills
\begin{align}
    \mathcal{T}^{K/K'}_{nm} (\Phi)&= \mathcal{T}^{K'/K}_{mn}(-\Phi).\label{eq.TOns}
\end{align}
This follows from Eq.~\eqref{eq.spintrans}. Equation \eqref{eq.TOns} implies the typical reciprocity relation \cite{datta_1995}
\begin{equation}
    G_{RL}(\Phi) = G_{LR} (-\Phi).
\end{equation}

\begin{figure}
	\includegraphics[width=0.48\textwidth]{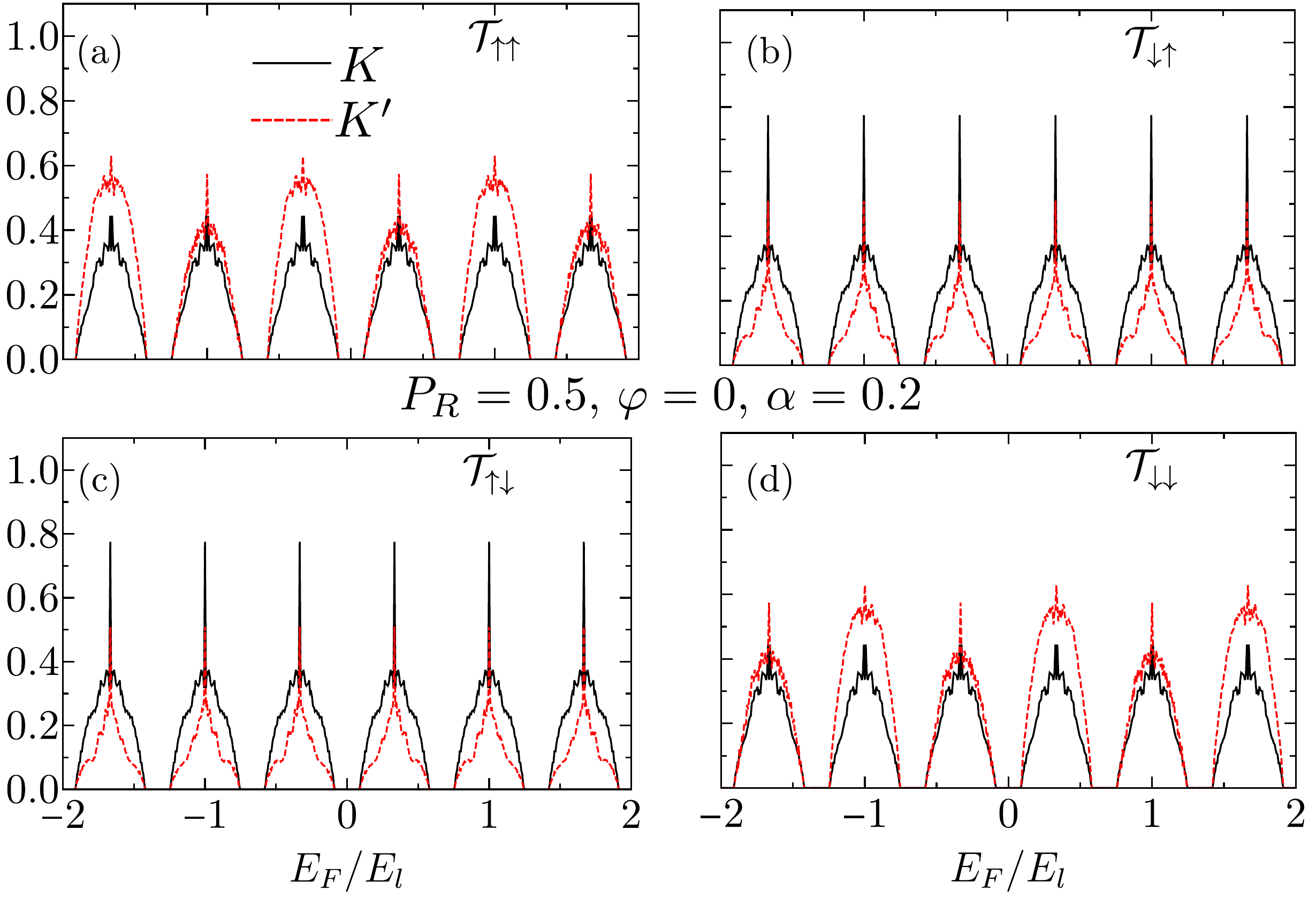}
	\caption{Spin-resolved transmission function from left to right of a network strip with length $L=10l$ for $P_R=0.5$, $\varphi=0$, $\alpha=0.2$  as a function of $E_F$ in units of $E_l=\pi \hbar v_F/l$. We show the transmission function $\mathcal{T}^K_{R\sigma, L\sigma'}$ for both valleys, where $\sigma'$ is the incoming and $\sigma$ is the outgoing spin.} \label{fig.CondSS}
\end{figure}
We show in Fig.~\ref{fig.CondSS} the transmission for $\Phi=0$ for a certain spin species calculated with the transmission function $\mathcal{T}^{K/K'}_{R\sigma,L \sigma'}$ as indicated in the inset of each panel. To simplify the notation we omit the index $R,L$, so that the total transmission function is given by
\begin{equation}
    \mathcal{T}= \sum_{\nu=K,K'}\left(\mathcal{T}^\nu_{\uparrow \uparrow} + \mathcal{T}^\nu_{\uparrow \downarrow}+\mathcal{T}^\nu_{\downarrow \downarrow}+\mathcal{T}^\nu_{\downarrow \uparrow}\right).
\end{equation}

First of all, we notice the finite deviations between $\mathcal{T}_{\uparrow \uparrow}$   and $\mathcal{T}_{\downarrow \downarrow}$ within each valley and between the valleys. Due to the angle dependence of the spin-orbit coupling [see Eq.~\eqref{eq.MSO}] every path accumulates a different phase and they are therefore inequivalent. Thus, the $S$ matrix of the contracted triangle shows deviations between different spin species depending on the path, which leads to this difference. 
Furthermore, we see a series of small sharp peaks for the spin-flip conductance [see Figs.~\ref{fig.CondSS}(b) and \ref{fig.CondSS}(c)] as we discussed in Sec.~\ref{Sec.CondSO}. At these specific points, the resonance conditions in Eqs.~\eqref{eq.cond1} and \eqref{eq.cond2} are fulfilled, which lead to these sharp peaks. Remarkably, these peaks are dominant in the spin-flip transmission, which is natural considering that under the resonance condition the spin flips for each transition between different nodes. 

To make a quantitative prediction of how spin polarized the transmission is, we introduce the spin polarization $P_\beta$ along the $\beta$ axis as \cite{Zhai2005}
\begin{align}
	P_x+i P_y&= \frac{2}{\mathcal{T}} \sum_{\nu=K,K'}\sum_{i,j,\sigma'} t^{\nu}_{i \downarrow,j\sigma'}(t^{\nu}_{i\uparrow, j \sigma'})^*,\\
	P_z &= \frac{1}{\mathcal{T}}\sum_{\nu=K,K'}\left(\mathcal{T}^\nu_{\uparrow \uparrow} + \mathcal{T}^\nu_{\uparrow \downarrow}-\mathcal{T}^\nu_{\downarrow \downarrow}-\mathcal{T}^\nu_{\downarrow \uparrow}\right),
\end{align}
where $i,j$ denote the modes going from left to right and $\mathcal{T}$ is the full transmission function. We also define the total polarization as
\begin{equation}
    P=\sqrt{P_x^2+P_y^2+P_z^2}.
\end{equation}

\begin{figure}
	\includegraphics[width=0.48\textwidth]{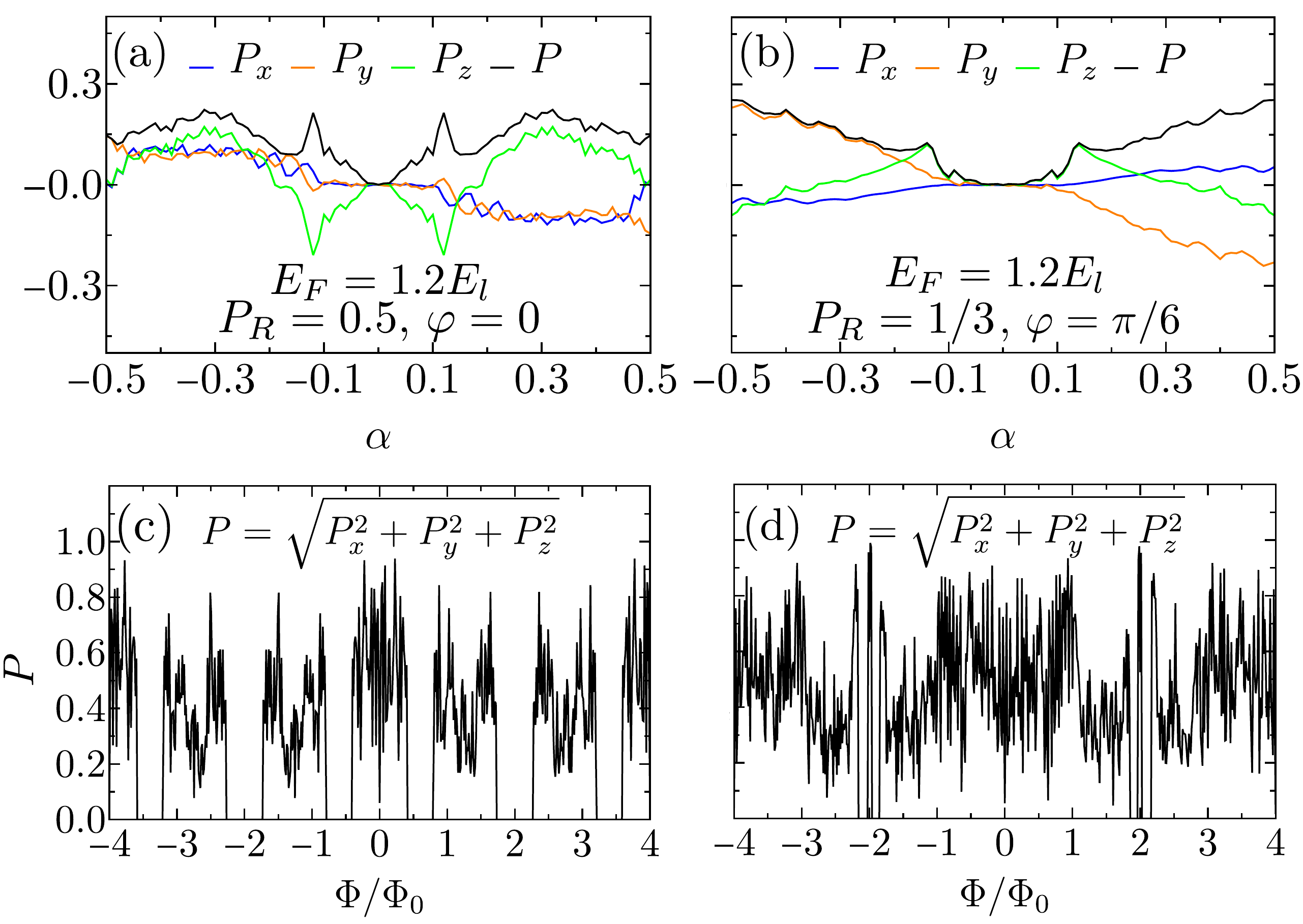}
	\caption{(a), (b) Polarization of the conductance of the network strip with length $L=10l$ for $P_R=0.5(1/3)$, $\varphi=0 (\pi/6)$ at $E_F=1.2E_l$ as a function of the spin-orbit coupling parameter $\alpha$ for $\Phi=0$. (c), (d) Polarization of the conductance of the network strip with length $L=10l$ for $P_R=0.5$, $\varphi=0 (0.2)$ at $E_F=E_l$ and $\alpha=0.2$ as a function of the flux $\Phi$ in units of $\Phi_0$.} \label{fig.PolSSAlpha}
\end{figure}

In Figs.~\ref{fig.PolSSAlpha}(a) and \ref{fig.PolSSAlpha}(b) we show results for the polarization of the conductance for different parameters for $\Phi=0$ as a function of the spin-orbit coupling parameter $\alpha$. For $\alpha\rightarrow 0$, we recover the spinless limit with zero polarization $P=P_\beta=0$. For larger values of $\alpha$, $P_\beta$ and $P$ vary non-monotonically respecting the symmetries $P_{x,y}(\alpha) = -P_{x,y}(-\alpha)$, $P_z(\alpha)=P_{z}(-\alpha)$. These symmetries are fulfilled, because the elements of the $S$ matrix fulfill
\begin{equation}
s^{K/K'}_{i\sigma, j \sigma'}(\alpha)=\sigma \sigma' s^{K/K'}_{i\sigma, j \sigma'}(-\alpha).
\end{equation}
The symmetries for the polarization as a function of spin-orbit coupling $\alpha$ hold within each valley for arbitrary $\Phi$.

For the sake of completeness, we show in Figs.~\ref{fig.PolSSAlpha}(c) and \ref{fig.PolSSAlpha}(d) the polarization at finite magnetic field as a function of the flux $\Phi$. The polarization depends on all parameters $P_R,\varphi,E_F$, and $\Phi$ in a non-monotonic and noisy fashion. Nevertheless, $P$ is still periodic in $8\Phi_0$. We note that the polarization in the other directions shows the same noisy behavior.

\section{Conclusions} \label{Conclusion}

In this paper, we have constructed a phenomenological scattering kagome network model based on the symmetries of the system. By combining $S$ matrices of single triangles, we have reduced the kagome network to a triangular network with an energy-dependent $S$ matrix. We have used this model to study the band structure and magnetotransport in different limits of the parameter regime finding Aharonov-Bohm resonances at finite temperature for integer values of the flux. This is in agreement with previous qualitative studies and generalizes former perturbation magnetotransport analysis \cite{Vladimir2022Kagome,devries2023kagome}.

Furthermore, motivated by the presence of Rashba spin-orbit coupling in graphene systems due to, for instance, proximity of transition metal dichalcogenides \cite{Avsar_2014,Gmitra2016,Garcia_2017,Wang2019,F_l_p_2021}, we have investigated its interplay with the kagome lattice structure. We find a localization effect in the network bands and also in the conductance due to the presence of a finite spin-orbit coupling. 
Moreover, we find conductance resonances that reflect the geometry of our system. These resonances occur when a spin-flip process takes place during the propagation between two nodes. In addition, this condition leads to the reduction of the periodicity in the magnetoconductance because an incoming electron with a certain spin needs to do two round trips around a triangle to go back to the same state instead of one. 

Lastly, we have studied the spin polarization of the current, finding a finite spin polarization in the presence of spin-orbit coupling due to interfering network paths with different phases induced by the angle dependence of the spin-orbit coupling. We observe numerically that the polarization varies in a noisy fashion as a function of all parameters in the model, reaching highly polarized values up to $P\approx 0.9$. 

Regarding experimental implications resulting from our work, it might be interesting to analyze a graphene layer under a periodic strain field \cite{DeBeule22Kagome} on a substrate that induces spin-orbit coupling, like $\text{WSe}_2$ \cite{Sun2023}. In such a system, the effects of the valley chiral kagome network such as Aharonov-Bohm conductance resonances and the spin-orbit effects should be both visible in transport measurements.

\subsection*{Acknowledgments}
We thank C. De Beule for fruitful discussions. We gratefully acknowledge the support of the Braunschweig International Graduate School of Metrology B-IGSM and the DFG Research Training Group 1952 Metrology for Complex Nanosystems. F. D. and P. R. gratefully acknowledge funding by the Deutsche Forschungsgemeinschaft (DFG, German Research Foundation) within the framework of Germany’s Excellence Strategy – EXC-2123 Quantum - Frontiers Grant No. 390837967.


%

\newpage
\appendix

\newpage
\section{$S$ matrix of one triangle with magnetic field} \label{App.triangularalpha0}
\begin{figure*}
	\centering
	\includegraphics[width=0.5\textwidth]{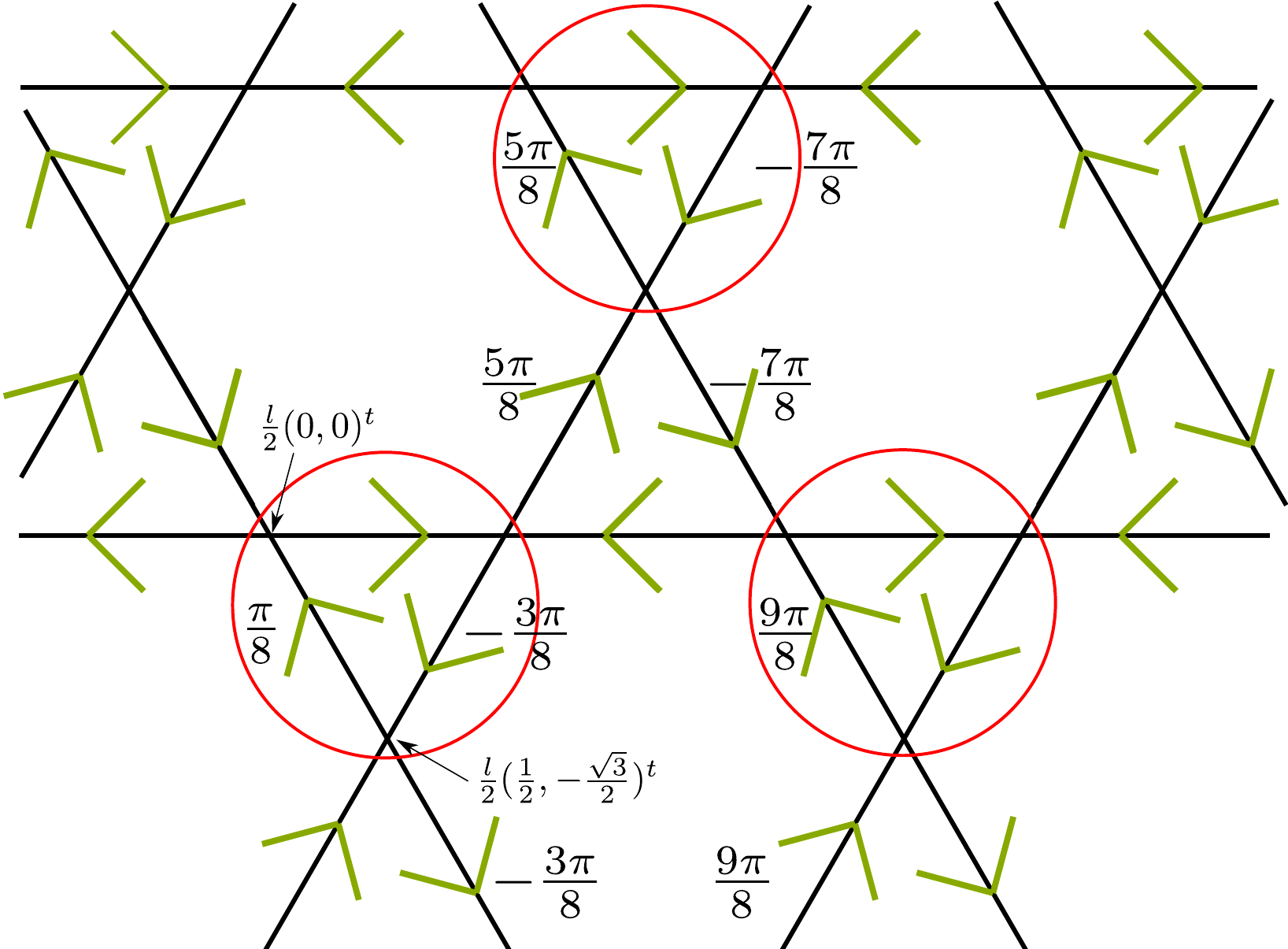}
	\caption{Kagome network. Red circles depict scattering matrices of single triangles, which we combine in the first step. Also, the Peierls phase accumulated between nodes is depicted here in the form $\phi_p=x \Phi/\Phi_0$, where $x$ can be found in the figure.} \label{fig.Flux}
\end{figure*}
Together with Eqs.~\eqref{eq.S1}--\eqref{eq.S3} and 
\begin{align}
\alpha_1&= \beta_1 \exp(i n \frac{\pi\Phi}{8\Phi_0})\exp(i \pi \epsilon/2),\\
\alpha_2&= \beta_2 \exp(i\pi \epsilon/2), \label{eq.dynApp1}\\
\alpha_3&= \beta_3 \exp(-i (n+2) \frac{\pi\Phi}{8\Phi_0})\exp(i \pi \epsilon/2),\label{eq.dynApp2}
\end{align}
where $\epsilon = E l/(\pi\hbar v_F)$ is the dynamical phase and $v_F$ is the Fermi velocity in graphene, we can eliminate the $\alpha_i$ and $\beta_i$ to find a $S$ matrix of the form
\begin{equation}
\begin{pmatrix}
b_1\\b_2\\b_3
\end{pmatrix} = S 	
\begin{pmatrix}
a_1\\a_2\\a_3
\end{pmatrix},
\end{equation}
with 
\begin{widetext}
	\begin{align}
	S_\triangle(\Phi)=&\frac{1}{e^{3i\varphi}-e^{-\frac{i \pi\Phi}{4\Phi_0}}e^{\frac{3i \pi \epsilon}{2}}P_R^{3/2}} \nonumber\\
	&\begin{pmatrix}
	-e^{2i \varphi}e^{-i (n+2) \frac{\pi}{8} \frac{\Phi}{\Phi_0}}e^{i \pi\epsilon}P_L \sqrt{P_R} &
	-e^{3i \varphi}e^{-i (n+2) \frac{\pi}{8} \frac{\Phi}{\Phi_0}}e^{\frac{i\pi\epsilon}{2}}P_L & 
	\left(e^{3i \varphi}-e^{-\frac{i \pi\Phi}{4\Phi_0}}e^{\frac{3i\pi\epsilon}{2}}\sqrt{P_R}\right)e^{i \varphi}\sqrt{P_R}\\
	\left(e^{3i \varphi}-e^{-\frac{i \pi\Phi}{4\Phi_0}}e^{\frac{3i\pi\epsilon}{2}}\sqrt{P_R}\right)e^{i \varphi}\sqrt{P_R} &
	-e^{2i \varphi}e^{-\frac{i \pi\Phi}{4\Phi_0}}e^{i \pi\epsilon}P_L \sqrt{P_R} & 
	-e^{3i \varphi}e^{i n \frac{\pi}{8} \frac{\Phi}{\Phi_0}}e^{\frac{i\pi\epsilon}{2}}P_L\\
	-e^{3i \varphi}e^{\frac{i\pi\epsilon}{2}}P_L
	& \left(e^{3i \varphi}-e^{-\frac{i \pi\Phi}{4\Phi_0}}e^{\frac{3i\pi\epsilon}{2}}\sqrt{P_R}\right)e^{i \varphi}\sqrt{P_R} & -e^{2i \varphi}e^{\frac{in \pi\Phi}{8\Phi_0}}e^{i \pi\epsilon}P_L \sqrt{P_R}
	\end{pmatrix}, \label{eq.Flux} 
	\end{align}
\end{widetext}
where $P_L = 1-P_R$. The parameter $n$ is an integer that is necessary for the conductance calculations and accounts for the position-dependent Peierls phase (see Fig.~\ref{fig.Flux}). For the magnetoconductance we have used the gauge $\mathbf{A}(x)=Bx \mathbf{e}_y$, where $\vec{e}_y$ is the standard Cartesian basis vector in the $y$ direction. Note that this matrix is $C_3$ symmetric for $\Phi=0$ as expected. The $S$ matrix for the other valley is given by $S_{K'}(\Phi)=S^t_\triangle(-\Phi)$.

\section{Calculation of Network bands with spin-orbit coupling}\label{App.Bands}

The $S$ matrix relating the incoming $a_{mn}$ with the outgoing $b_{mn}$ modes at a given node $(m,n)$ at position $m \bm{l}_1+n \bm{l}_2$ ($m,n\in \mathds{Z}$) is given by
\begin{align}
b_{mn,\sigma}=S a_{mn,\sigma}
\end{align} 
 with $S=S_\sigma\otimes\mathds{1}_{2\cross2}$ and
\begin{align}
a_{mn}=&(a_{1mn,\uparrow},a_{1mn\downarrow},\alpha_{1,\uparrow},\alpha_{1,\downarrow},a_{2mn,\uparrow},a_{2mn,\downarrow},\nonumber\\
&\alpha_{2,\uparrow},\alpha_{2,\downarrow},a_{3mn,\uparrow},a_{3mn,\downarrow},\alpha_{3,\uparrow},\alpha_{3,\downarrow})^t,\\
b_{mn}=&(b_{2mn,\uparrow},b_{2mn,\downarrow},\beta_{2,\uparrow},\beta_{2,\downarrow},b_{3mn,\uparrow},b_{3mn,\downarrow},\nonumber\\
&\beta_{3,\uparrow},\beta_{3,\downarrow},b_{1mn,\uparrow},b_{1mn,\downarrow},\beta_{1,\uparrow},\beta_{1,\downarrow})^t.
\end{align}
 Note the absence of $mn$ indices in the inner triangle scattering amplitudes $\alpha_{i,\sigma},\beta_{i,\sigma}$, which are equal at every node. 
 
 Using the Bloch's theorem, we relate the outgoing states of different unit cells by means of 
 \begin{equation} \label{eq.blochalpha}
\begin{pmatrix}  b_{1m+1n,\uparrow} \\b_{1m+1n,\downarrow} \\ \beta_{1,\uparrow} \\ \beta_{1,\downarrow} \\ b_{2m-1n-1,\uparrow} \\b_{2m-1n-1,\downarrow} \\ \beta_{2,\uparrow} \\ \beta_{2,\downarrow} \\ b_{3mn+1,\uparrow} \\b_{3mn+1,\downarrow} \\ \beta_{3,\uparrow} \\ \beta_{3,\downarrow} \end{pmatrix}=  \mathcal M(\bm k)  \begin{pmatrix} b_{2mn,\uparrow} \\b_{2mn,\downarrow} \\ \beta_{2,\uparrow} \\ \beta_{2,\downarrow} \\ b_{3mn,\uparrow} \\b_{3mn,\downarrow} \\ \beta_{3,\uparrow} \\ \beta_{3,\downarrow} \\ b_{1mn,\uparrow} \\b_{1mn,\downarrow} \\ \beta_{1,\uparrow} \\ \beta_{1,\downarrow}  \end{pmatrix},
\end{equation}
with $\mathcal M(k)= \mathcal{M}_\sigma(k) \otimes \mathds{1}_{2\cross2}$ and
\begin{equation}
\mathcal M_\sigma(k)= \begin{pmatrix}
0 & 0 & 0 & 0 &e^{i k_1} & 0\\
0 & 0 & 0 & 0 & 0 &1\\
e^{i k_3} & 0 & 0 & 0 &0 & 0\\
0 & 1 & 0 & 0 & 0& 0\\
0 & 0 & e^{i k_2} & 0 &0 & 0\\
0 & 0 & 0 & 1 &0 & 0\\
\end{pmatrix}.
\end{equation}

Using Eqs.~\eqref{eq.dyn1}--\eqref{eq.dyn3} and \eqref{eq.dyn4}--\eqref{eq.dyn6}, we relate

\begin{equation} \label{eq.epsalpha}
\begin{pmatrix} a_{1mn,\uparrow} \\a_{1mn,\downarrow} \\ \alpha_{1,\uparrow} \\ \alpha_{1,\downarrow} \\ a_{2mn,\uparrow} \\a_{2mn,\downarrow} \\ \alpha_{2,\uparrow} \\ \alpha_{2,\downarrow} \\ a_{3mn,\uparrow} \\a_{3mn,\downarrow} \\ \alpha_{3,\uparrow} \\ \alpha_{3,\downarrow}  \end{pmatrix}=D(\alpha,\epsilon)\begin{pmatrix}  b_{1m+1n,\uparrow} \\b_{1m+1n,\downarrow} \\ \beta_{1,\uparrow} \\ \beta_{1,\downarrow} \\ b_{2m-1n-1,\uparrow} \\b_{2m-1n-1,\downarrow} \\ \beta_{2,\uparrow} \\ \beta_{2,\downarrow} \\ b_{3mn+1,\uparrow} \\b_{3mn+1,\downarrow} \\ \beta_{3,\uparrow} \\ \beta_{3,\downarrow} \end{pmatrix}  ,
\end{equation}
with the transfer matrix
	\begin{align}
	D(\alpha,\epsilon)= \text{diag} [ M(-\pi/3),  &M(2\pi/3), M(\pi),M(0),\nonumber \\
 &M(\pi/3),M(-2\pi/3)]
  \label{App.D}
	\end{align}
which reduces to $D(0,\epsilon)=e^{i\pi \epsilon/2}\mathds{1}_{12\cross 12}$ in the absence of spin-orbit coupling. 

Now, we combine Eq.~\eqref{eq.blochalpha} with~\eqref{eq.epsalpha} to obtain
\begin{equation}
D(\alpha,\epsilon)^{-1} a_{\bm k} =   \mathcal M(\bm k)  b_{\bm k}=\mathcal M(\bm k)  S a_{\bm k},
\end{equation}
from which we obtain the equation to compute the network energy bands,
yielding the condition to obtain the energy spectrum of the network, that is,
\begin{align}
\det[\mathcal M(\bm k) S- D^{-1}] =0.
\end{align}

\section{Combining procedure} \label{App.Combining}

\begin{figure}
	\includegraphics[width=0.4\textwidth]{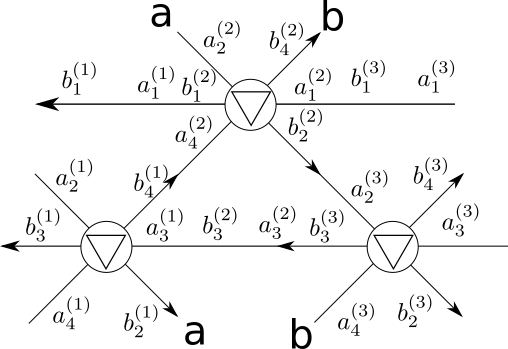}
	\caption{Minimal element of the network that is repeatedly combined to calculate the transmission function of the network strip. The spin index is omitted here.} \label{fig.network}
\end{figure}

Here, we explain how we have calculated the transmission function $\mathcal{T}(E)$ for our transport calculations in Eq.~\eqref{eq.Cond} by the combination of scattering matrices. We show the general idea how to combine the first two scattering matrices. Finally, we explain the recursive loop that we have implemented for our calculations.
One part of the network is shown in Fig~\ref{fig.network}. The first $S$ matrix is given by 
\begin{equation}
S^{(1)} =\begin{pmatrix}
\mathds{1} & 0\\
0& 	S\\ 
\end{pmatrix}
\end{equation}
in the basis $(b_{1,\uparrow}^{(1)},b_{1,\downarrow}^{(1)},b_{2,\uparrow}^{(1)},b_{2,\downarrow}^{(1)},b_{3,\uparrow}^{(1)},b_{3,\downarrow}^{(1)},b_{4,\uparrow}^{(1)},b_{4,\downarrow}^{(1)})^t=S^{(1)} (a_{1,\uparrow}^{(1)},a_{1,\downarrow}^{(1)},a_{2,\uparrow}^{(1)},a_{2,\downarrow}^{(1)},a_{3,\uparrow}^{(1)},a_{3,\downarrow}^{(1)},a_{4,\uparrow}^{(1)},a_{4,\downarrow}^{(1)})^t$, where $S$ is the $S$ matrix $S_\triangle(\Phi)\otimes \mathds{1}_{2\cross2}$ [see Eq.~\eqref{eq.Flux}] for $\alpha=0$, or for $\alpha\neq0$ see Eq.~\eqref{eq.SmatrixSO2}.

The second $S$ matrix of the network can be calculated from $S^{(1)}$ by interchanging the first and the third modes, i.e.,
\begin{equation}
S^{(2)}=R_{13} S_1 R_{13}, \, R_{13}= \begin{pmatrix}
0 & 0 & \mathds{1} & 0\\
0 & \mathds{1} & 0 & 0\\
\mathds{1} & 0 & 0 & 0\\
0 & 0 & 0 & \mathds{1}\\
\end{pmatrix}.
\end{equation}
Now we need to combine these two matrices. Before we look into that, we change the basis a little bit to make the process of combining easier. We rewrite $S^{(1)}$, so that it fulfills
\begin{align}
\begin{pmatrix}
b_L^{{(1)}}\\
b_R^{{(1)}}\\
\end{pmatrix} = \begin{pmatrix}
r_L^{{(1)}} & t_{LR}^{{(1)}}\\
t_{RL}^{{(1)}} & r_R^{{(1)}}
\end{pmatrix} \begin{pmatrix}
a_L^{(1)}\\
a_R^{(1)}\\
\end{pmatrix}.  \label{eq.S1n}
\end{align}
To bring these notations together we define 
\begin{align}
b_L^{(1)}&:= \begin{pmatrix}
b_{1,\uparrow}^{(1)},b_{1,\downarrow}^{(1)}, b_{3,\uparrow}^{(1)}, b_{3,\downarrow}^{(1)}
\end{pmatrix}^t, \\
b_R^{(1)}&:= \begin{pmatrix}
b_{2,\uparrow}^{(1)},b_{2,\downarrow}^{(1)}, b_{4,\uparrow}^{(1)}, b_{4,\downarrow}^{(1)}
\end{pmatrix}^t,\\
a_L^{(1)}&:= \begin{pmatrix}
a_{4,\uparrow}^{(1)},a_{4,\downarrow}^{(1)}, a_{2,\uparrow}^{(1)}, a_{2,\downarrow}^{(1)}
\end{pmatrix}^t,\\
a_R^{(1)}&:= \begin{pmatrix}
a_{1,\uparrow}^{(1)},a_{1,\downarrow}^{(1)}, a_{3,\uparrow}^{(1)}, a_{3,\downarrow}^{(1)}
\end{pmatrix}^t.
\end{align}
With that we bring the $S$ matrix into a block structure with the submatrices $r_{i}$ that contain the reflection processes to the direction $i=L,R$, and $t_{ij}$, that contains the transmission processes from $j=L,R$ to $i=L,R$, where $L$ means left and $R$ means right. Note that $r_{i}$ and $t_{ij}$ are $4\times 4$ matrices. We can write it in the following way:
\begin{align}
r_L^{(1)}&=\begin{pmatrix}
0 & 0\\
s_{l,2} & s_{r,2}
\end{pmatrix},\,
t_{RL}^{(1)}=\begin{pmatrix}
s_{r,1} & s_{f,3}\\
s_{f,2} & s_{l,3}
\end{pmatrix},\\
r_R^{(1)}&=\begin{pmatrix}
0 & s_{l,1}\\
0 & s_{r,3}
\end{pmatrix},\,
t_{LR}^{(1)}=\begin{pmatrix}
1 & 0\\
0 & s_{f,2}
\end{pmatrix},\\
s_{f,1} &= \begin{pmatrix}
s_{11} & s_{12}\\
s_{21} & s_{22}
\end{pmatrix},\, 
s_{f,2} = \begin{pmatrix}
s_{33} & s_{34}\\
s_{43} & s_{44}
\end{pmatrix},\, 
s_{f,3} = \begin{pmatrix}
s_{55} & s_{56}\\
s_{65} & s_{66}
\end{pmatrix},\\
s_{l,1} &= \begin{pmatrix}
s_{13} & s_{14}\\
s_{23} & s_{24}
\end{pmatrix},\,
s_{l,2} = \begin{pmatrix}
s_{35} & s_{36}\\
s_{45} & s_{46}
\end{pmatrix},\,
s_{l,3} = \begin{pmatrix}
s_{51} & s_{52}\\
s_{61} & s_{62}
\end{pmatrix},\\
s_{r,1} &= \begin{pmatrix}
s_{15} & s_{16}\\
s_{25} & s_{26}
\end{pmatrix},\,
s_{r,2} = \begin{pmatrix}
s_{31} & s_{32}\\
s_{41} & s_{42}
\end{pmatrix},\,
s_{r,3} = \begin{pmatrix}
s_{53} & s_{54}\\
s_{53} & s_{64}
\end{pmatrix},\,
\end{align}
where $s_{ij}$ is the matrix element of $S$ in Eq.~\eqref{eq.SmatrixSO2}.

The second $S$ matrix can be written in a similar way as
\begin{align}
\begin{pmatrix}
b_L^{{(2)}}\\
b_R^{{(2)}}\\
\end{pmatrix} = \begin{pmatrix}
r_L^{{(2)}} & t_{LR}^{{(2)}}\\
t_{RL}^{{(2)}} & r_R^{{(2)}}
\end{pmatrix} \begin{pmatrix}
a_L^{(2)}\\
a_R^{(2)}\\
\end{pmatrix}  \label{eq.S2n}
\end{align}
with
\begin{align}
b_L^{(2)}&:= \begin{pmatrix}
b_{1,\uparrow}^{(2)},b_{1,\downarrow}^{(2)}, b_{3,\uparrow}^{(2)},b_{3,\downarrow}^{(2)}
\end{pmatrix}^t,\\
b_R^{(2)}&:= \begin{pmatrix}
b_{4,\uparrow}^{(2)},b_{4,\downarrow}^{(2)}, b_{2,\uparrow}^{(2)},b_{2,\downarrow}^{(2)}
\end{pmatrix}^t,\\
a_L^{(2)}&:= \begin{pmatrix}
a_{2,\uparrow}^{(2)}, a_{2,\downarrow}^{(2)},a_{4,\uparrow}^{(2)},a_{4,\downarrow}^{(2)}
\end{pmatrix}^t,\\
a_R^{(2)}&:= \begin{pmatrix}
a_{1,\uparrow}^{(2)},a_{1,\downarrow}^{(2)}, a_{3,\uparrow}^{(2)},a_{3,\downarrow}^{(2)}
\end{pmatrix}^t,\\
r_L^{(2)}&=\begin{pmatrix}
e^{-i \sqrt{3}k \frac{l}{2}}s_{r,2} & s_{l,2}\\
0 & 0
\end{pmatrix},\\
t_{RL}^{(2)}&=\begin{pmatrix}
s_{l,3} & e^{i\sqrt{3}k\frac{l}{2}}s_{f,3}\\
e^{-i\sqrt{3}k\frac{l}{2}}s_{f,1} & s_{r,1}\\
\end{pmatrix},\\
r_{R}^{(2)}&=\begin{pmatrix}
e^{i\sqrt{3}k\frac{l}{2}}s_{r,3} & 0\\
  s_{l,1} & 0
\end{pmatrix},\\
t_{LR}^{(2)}&=\begin{pmatrix}
s_{f,2} & 0\\
0 & 1
\end{pmatrix}.
\end{align}
We have added in the second $S$ matrix also the transversal momentum \cite{DeBeule2020} $0\leq k<4\pi/\sqrt{3}l$. Due to the translational symmetry in the $y$ direction, the modes that leave and enter Fig.~\ref{fig.network} in the $y$ direction are related by Bloch's theorem. We integrate over the transversal momentum at the end.
Now we need to know how the incoming and outgoing modes are related. To do so we define 
\begin{equation}
W(\theta) := \exp\left[ i \pi  s_0\epsilon/4\right]\exp\left\{i \pi \alpha [s_x\sin(\theta)-s_y\cos(\theta)]/4\right\}.
\end{equation} Then it follows
\begin{align}
&a_R^{(1)} = \gamma b_L^{(2)}, \, a_L^{(2)} = \beta_1 b_R^{(1)}, \label{eq.conn}
\end{align}
with
\begin{align}
&\gamma =  \mathds{1}_{2\cross 2} \otimes W(\pi),\\
&\beta_{1} = \begin{pmatrix}
W^2(-\pi/3) & 0 \\
0 & W^2(\pi/3)
\end{pmatrix} * \left[P_1\otimes \mathds{1}_{2\cross 2}\right],\\
&\beta_{2} = \begin{pmatrix}
W^2(\pi/3) & 0 \\
0 & W^2(-\pi/3)
\end{pmatrix} * \left[P_2\otimes \mathds{1}_{2\cross 2}\right],\\
&P_1(\Phi,n) = \begin{pmatrix}
 e^{-i(n-2)\frac{\pi}{8}\frac{\Phi}{\Phi_0}} & 0 \\
0 &  e^{in\frac{\pi}{8}\frac{\Phi}{\Phi_0}}\\
\end{pmatrix},\\
&P_2(\Phi,n) = \begin{pmatrix}
 e^{in\frac{\pi}{8}\frac{\Phi}{\Phi_0}} & 0 \\
0 &  e^{-i(n-2)\frac{\pi}{8}\frac{\Phi}{\Phi_0}}
\end{pmatrix}.
\end{align}
The matrix $\beta_n$ contains the dynamical phase and the Peierls phase due to the magnetic field of a mode traversing from one node to the next. The matrix $\gamma$ contains the phase $\epsilon$ and the dynamical phase of a mode traversing in $x$ direction. Therefore, it does not accumulate a Peierls phase due to the used gauge $\mathbf{A}=Bx \mathbf{e}_y$. The parameter $n\in 4\mathds{N}+1$ will be counted up for every combining step. The dynamical phase is influenced by the spin-orbit coupling $\alpha$.
To combine these matrices we can write 

\begin{align}
\mathbf{r}_L^{(1+2)}&=r_L^{(1)}+t_{LR}^{(1)}\gamma Q_2 r_L^{(2)}\beta_1 t_{RL}^{(1)}, \label{eq.step11}\\
\mathbf{r}_R^{(1+2)}&=r_R^{(2)}+t_{RL}^{(2)}\beta_1 Q_1 r_{R}^{(1)}\gamma t_{LR}^{(2)},\\
\mathbf{t}_{LR}^{(1+2)}&=t_{LR}^{(1)}\gamma Q_2 t_{LR}^{(2)},\\
\mathbf{t}_{RL}^{(1+2)}&=t_{RL}^{(2)} \beta_1 Q_1 t_{RL}^{(1)},\\
Q^{(1+2)}_1&=\left(1-r_R^{(1)}\gamma r_L^{(2)}\beta_1\right)^{-1},\\
Q^{(1+2)}_2&=\left(1-r_L^{(2)}\beta_1 r_R^{(1)}\gamma\right)^{-1}. \label{eq.step14}
\end{align}
The next $S$ matrix is the same as the first one. To combine this third $S$ matrix with the already calculated one we can write
\begin{align}
\mathbf{r}_{L}^{(1+2+3)}&=\mathbf{r}_L^{(1+2)}+\mathbf{t}_{LR}^{(1+2)}\gamma Q^{(1+2+3)}_2r_L^{(1)}\beta_2\mathbf{t}_{RL}^{(1+2)},\\
\mathbf{r}_R^{(1+2+3)}&=r_R^{(1)}+t_{RL}^{(1)}\beta_2 Q^{(1+2+3)}_1 \mathbf{r}_{R}^{(1+2)}\gamma t_{LR}^{(1)},\\
\mathbf{t}_{LR}^{(1+2+3)}&=t_{LR}^{(1)}\gamma Q^{(1+2+3)}_2 \mathbf{t}_{LR}^{(1+2)},\\
\mathbf{t}_{RL}^{(1+2+3)}&=\mathbf{t}_{RL}^{(1+2)}\beta_2 Q^{(1+2+3)}_1t_{RL}^{(1)},\\
Q^{(1+2+3)}_1&=\left(1-\mathbf{r}_{R}^{(1+2)}\gamma r_L^{(1)}\beta_2\right)^{-1},\\
Q^{(1+2+3)}_2&=\left(1-r_L^{(1)}\beta_2\mathbf{r}_R^{(1+2)}\gamma\right)^{-1}.
\end{align}
By replacing 
\begin{align}
&r_L^{(1)} \rightarrow \mathbf{r}_L^{(1+2+3)}, \, t_{RL}^{(1)} \rightarrow \mathbf{t}_{RL}^{(1+2+3)},\\
&r_{R}^{(1)} \rightarrow \mathbf{r}_{R}^{(1+2+3)}, \, t_{LR}^{(1)} \rightarrow \mathbf{t}_{LR}^{(1+2+3)}
\end{align}
in Eqs.~\eqref{eq.step11}--\eqref{eq.step14} we can loop this procedure to calculate the $S$ matrix of the network.

With the $S$ matrix of the network we can then calculate the conductance. 
The transmission function per unit cell can be calculated from the $S$ matrix for one valley by means of
\begin{equation}
\mathcal{T}^K_{RL}(E) = (\sqrt{3}l/4\pi)\int_{0}^{4\pi/\sqrt{3}l} dk \text{Tr}[t_{RL}^\dagger t_{RL}], \label{eq.Tfunction}
\end{equation}
where $t_{RL}$ are the transmission matrix elements of the $S$ matrix from the left side to the ride side of the strip. Note that the transmission function per unit cell of one valley is not necessarily the same for the other valley in the presence of a magnetic field. One can show that for finite flux $\Phi$ the transmission function $\mathcal{T}_{nm}$ of the valley $K$ and $K'$ are related by
\begin{equation}
\mathcal{T}^K_{nm}(\Phi)=\mathcal{T}^{K'}_{mn}(-\Phi),
\end{equation}
with $m,n=L,R$.
Also in the presence of spin-orbit coupling, the angle $\theta$ changes to $\theta+\pi$ in the other valley, because the modes traverse in the other direction. The total transmission function is
\begin{equation}
\mathcal{T}_{nm}(E) = \mathcal{T}^K_{nm}+\mathcal{T}^{K'}_{nm}.
\end{equation}

\end{document}